\documentstyle[10pt,makeidx]{report}                 % VUB

\makeindex

\begin{document}
\pagenumbering{arabic}

%\chapter*{Painlev\'e analysis for nonlinear partial differential equations}
\begin{center}
    {\bf PAINLEV\'E ANALYSIS FOR NONLINEAR PARTIAL DIFFERENTIAL EQUATIONS}
\end{center}
\vskip 0.5 truecm

%\chapterauthors{Micheline Musette}

{\bf Micheline Musette}

Dienst Theoretische Natuurkunde,
Vrije Universiteit Brussel, Pleinlaan 2

B--1050 Brussel

\vskip 0.5 truecm

\begin{center}
{\it Proceedings of the Carg\`ese school (3--22 June 1996)}
\\
{\it La propri\'et\'e de Painlev\'e, un si\`ecle apr\`es}
\\
{\it The Painlev\'e property, one century later}
\\
%\date
\end{center}

% Forcer la pr\'esence du num\'ero de chapitre dans \ref{sectionlabel} MODIF
% Forcer les \subsubsection \`a ``se voir'' dans le texte              MODIF
%\renewcommand{\subsubsection}{\subsection} % Transformer en %         MODIF
%\setcounter{secnumdepth}{3} % number sections, subsections and subsubsections
%\setcounter{tocdepth}{3}    % and put them in table of contents

% ********************************************************** MACROS.TEX, start
\def\vstrut{\phantom{^{\displaystyle |}}}

\def \CRAS {C.~R.~Acad.~Sc.~Paris}
% ----------------------------------------- Language independent abbreviations
\def \Plv{Pain\-le\-v\'e}
\def \KdV {Kor\-te\-weg-de Vries}
% ------------------------------------------------------ English abbreviations
\def \pde {partial differential equation}
\def \pdes{partial differential equations}
% ----------------------------------------------------- Mathematical functions
\def \D {\hbox{d}}
\def \petD {\hbox{\scriptsize d}}
\def \Log {\mathop{\rm Log}\nolimits}
\def \sinh{\mathop{\rm sinh}\nolimits}
\def \cotanh{\mathop{\rm cotanh}\nolimits}
\def \sech{\mathop{\rm sech}\nolimits}
\def \sn  {\mathop{\rm sn}\nolimits}
\def \cn  {\mathop{\rm cn}\nolimits}
\def \Re  {\mathop{\rm Re}\nolimits}
\def \Im  {\mathop{\rm Im}\nolimits}
\def \arg {\mathop{\rm arg}\nolimits}
\def \mod#1{\vert #1 \vert}
% ------------------------------------------- Complex Ginzburg-Landau equation
\def\GL{Ginz\-burg-Lan\-dau}
\def\GLA{A}
\def\GLB{B}
\def\pcarre{{\vert p \vert}^2}
\def\pqr{D_r}
\def\pqi{D_i}
\def \expon {\alpha} % \Theta_{01}
% *********************************************************** MACROS.TEX, end
%

\vskip 0.5 truecm
{\it Abstract}.
The Painlev\'e analysis introduced by Weiss, Tabor and Carne\-vale (WTC) 
in 1983
for nonlinear partial differential equations (PDE's) is an extension of the 
method initiated by Painlev\'e and Gambier at the beginning of this century
for the classification of algebraic nonlinear differential equations (ODE's)
without movable critical points. 
In these lectures we explain the WTC method in its invariant version
introduced by Conte in 1989 and its application to solitonic equations in
order to find algorithmically their associated B\"acklund transformation. 
A lot of remarkable  properties are shared by these so-called ``integrable''
equations but they are generically no more valid for equations modelising 
physical phenomema. 
Belonging to this second class, some equations called ``partially integrable''
sometimes keep remnants of integrability. 
In that case, the singularity analysis may also be useful for building 
closed form analytic solutions, which necessarily % Conte
agree with the singularity structure of the equations. 
We display the privileged role played by the Riccati equation and systems of
Riccati equations which are linearisable,
as well as the importance of the Weierstrass elliptic function,
for building solitary waves or more elaborate solutions.  

\vfill

{\it The Painlev\'e property, one century later},
ed.~R.~Conte, 
CRM series in mathematical physics (Springer--Verlag, Berlin, 1998)
\hfill 31~March~1998

\hfill \hskip 1 truecm solv-int/9804003

\tableofcontents

% ==========================================================================
\chapter{Introduction}
\indent

%Aim of my lecture.

%Two main classes of PDEs.

During the past thirty years, the interest for nonlinear phenomena has been
growing in different fields of modern physics, such as optics, 
fluid dynamics, condensed matter, elementary particle physics, 
statistical mechanics, astrophysics.
Although the manifestation of those phenomena varies according to
the different fields, they present a common feature in their mathematical
description. 
The link comes from their description by
{\it nonlinear evolution equations} (i.e.~PDEs) 
whose solutions represent the propagation of waves with a permanent profile. 
Moreover, the analytical methods for solving them are directly inspired by
the works of the famous mathematicians L.~Fuchs, 
H.~Poincar\'e and P.~Painlev\'e, as explained in Conte contribution,
whose content is assumed known.

The propagation of a bell-shaped solitary wave \index{solitary wave} on water
has been approximately explained by the mathematical physicists
J.~Boussinesq \cite{Boussinesq} and Lord Rayleigh \cite{Rayleigh}
only thirty years after its experimental discovery by Scott Russell in 1844
\cite{Russell}.
The full explanation was later given in 1895  by Korteweg and de Vries 
(KdV \cite{KdV1895}) who
derived the nonlinear dispersive equation
\index{KdV equation}
\begin{equation}
\label{eqKdVcons}
 u_t + u_{xxx} + 3 (u^2)_x = 0, 
\end{equation}
possessing the two-parameter $(k,\tau)$ exact solution
\begin{equation}
\label{eqKdVsw}
u^{\rm sw}(k,\theta)
={k^2 \over 2} \sech^2 {\theta \over 2},\ \sech={1 \over \cosh},\ 
\theta=k \xi + \tau,
\xi=x-ct,\ c=k^2. 
\end{equation}

The name soliton was introduced by Zabusky and Kruskal in 1965 \cite{ZaKru}
when they solved the initial value problem for KdV equation (\ref{eqKdVcons})
and discovered solutions describing the elastic collision of
several waves (\ref{eqKdVsw}).
%For the reader not familiar with the soliton theory,
%we recall some usual definitions
%\begin{itemize}
%\item a \underbar{travelling wave} of a PDE is a solution of the ODE
%obtained by the
%       reduction $(x,t) \to \xi=x-ct,$ % = permanent profile
% \item a \underbar{solitary wave}\index{solitary wave} is a travelling wave 
% going to zero,
%as well as all its derivatives, as $\mod{\xi} \to + \infty$,
% \item a \underbar{soliton} is a solitary wave which, 
%after interaction with other solitons,
%conserves its profile and is only modified by a constant phase shift.
%The PDE possesses an exact solution $u_N(x,t),\ N \ge 2$, 
% (not yet written here)
%called \underbar{$N$-soliton} solution, if it satisfies for large time in 
% the past and in the future the superposition principle~:
%\begin{eqnarray}
%\lim_{t \rightarrow - \infty} u_N(x,t) 
%& \sim &
%\sum_{i=1}^N u^{\rm sw}(k_i,\theta_i)
%\\
%\lim_{t \rightarrow + \infty} u_N(x,t)
%& \sim &
%\sum_{i=1}^N u^{\rm sw}(k_i,\theta_i + \delta_i)
%\end{eqnarray}
%in which the velocities $c_i$ are all different.
%\end{itemize}

In these lectures we shall restrict our study by the method of singularities
to nonlinear evolution equations possessing two different levels of 
integrability~: 
{\it complete integrability} or {\it partial integrability}, 
including some chaotic PDE's
which possess explicit analytic solutions in very special circumstances.

{\it Complete integrability} means that
\index{Painlev\'e property for PDEs}
\index{Painlev\'e test for PDEs}
\begin{itemize}
 \item either the nonlinear {\pde} can be related to a linear {\pde} by an 
explicit transformation,
 \item or the equation passes the Painlev\'e test and possesses the {\Plv} 
property (PP) for PDEs, 
i.e.~: 
{\it firstly}, on every noncharacteristic manifold its general solution
has no movable critical singularities in the complex plane of an arbitrary
function $\varphi(x,t)$; 
{\it secondly}, the PDE possesses an auto-B\"acklund transformation
\index{B\"acklund transformation} 
or is related by a B\"acklund transformation
to another PDE possessing the PP (PDEs passing the `` weak Painlev\'e'' test
\cite{AblowitzClarkson} and related by a hodograph transformation
\cite{CFA1989} to another equation possessing the PP are outside the scope
of these lectures),  
 \item or the equation possesses 
solitary waves
\index{solitary wave}, 
$N-$soliton solutions for arbitrary $N$, 
an infinite number of conservation laws, 
bi-Hamiltonian structures,
infinite-dimensional Lie algebras,
\dots
 \item or the equation satisfies the Ablowitz-Ramani-Segur (ARS) 
\cite{ARS1978,ARS1980}
conjecture on the relationship of \underbar{all} its reductions to ODE's 
without  movable critical points. 
\end{itemize}
More explicit definitions concerning the properties of this first class of 
equations, 
as well as classical examples will be given in chapter \ref{chap2}.

{\it Partial integrability} means that some above listed properties 
are not satisfied
(in particular the Painlev\'e test may be satisfied only with some constraints
on the function $\varphi$,
or may never be satisfied whatever be $\varphi$,
and the ARS conjecture is no more valid)
but the equation possesses explicit analytic solutions like for instance~: 
degenerate solitary waves\index{solitary wave},
$N-$shock solutions,
$N-$soliton solutions with $N$ bounded \cite{BBM,FokasLiu1994,LiuFokas1996}, 
or % ? grammaire
retains some pieces of integrability like 
degenerate B\"acklund transformations,
\index{B\"acklund transformation}
a finite number of conservation laws \cite{BBM,Olver1979,Harris1996}.
For equations belonging to this second class,
methods for finding particular solutions,
which must agree with the singularity structure of the equation,
will be developed in chapter \ref{chap4}.

Chapter \ref{chap3} contains the main subject of our lectures~:
it is devoted to the WTC \cite{WTC} method and its extensions in the 
invariant version introduced by Conte \cite{Conte1989},
for finding algorithmically the auto-B\"acklund transformation
\index{B\"acklund transformation} 
of integrable nonlinear PDEs.

% ==========================================================================
\chapter{Integrable equations}
\label{chap2}
\indent

We present here a few classical examples of nonlinear {\pdes} either
explicitly related to linear {\pdes} or characterised by the properties of
complete integrability mentioned in the previous chapter.

% ---------------------------------------------------------------------------
\section{Integration by direct linearisation}
\indent

Some equations can be linearised by an explicit transformation~:

{\it Example} 1. The {\bf Burgers} equation
\index{Burgers equation}
\begin{equation}
\label{eqBurgers}
 u_t + (u_x + u^2)_x = 0
\end{equation}
is linearised into the heat equation \cite{Forsyth1906}
% Florin 1948, Hopf 1950, Cole 1951
\begin{equation}
\label{eqBurgers1}
u=(\Log \varphi)_x,\ 
((\varphi_t + \varphi_{xx}) / \varphi)_x=0.
\end{equation}

%\begin{equation}
% u=(\Log \varphi)_x,\ \varphi_t + \varphi_{xx} = F(t) \varphi,\
%F \hbox{ arbitrary function},
%\end{equation}
%an equation equivalent to the heat equation by a gauge transformation
%\begin{equation}
% \varphi = \Phi e^{\int^t F\ \D t},\
%\Phi_t + \Phi_{xx} = 0.
%\end{equation}

{\it Example} 2. The generalised {\bf Eckhaus} equation 
\cite{CalogerodeLillo,ClarksonCosgrove1987,Kundu1988}
\index{Eckhaus equation} % Kundu-Eckhaus?
\begin{equation}
 i u_t + u_{xx} 
+(\beta^2 {\mod{u}}^4 +2 \beta e^{i \gamma}({\mod{u}}^2)_x)u=0,\
(\beta, \gamma) \in {\cal R},\ 
\end{equation}
is linearisable into the Schr\"odinger equation for $\beta \cos \gamma \not=0$
\begin{eqnarray}
\hspace{-5.5truemm} i \nu_t +  \nu_{xx} = 0,\
u& = &\sqrt{{1 \over 2 \beta \cos \gamma}} {\nu \over \sqrt\varphi} 
e^{-(i/2) \tan\gamma\Log \varphi},\mbox{ with }\varphi_x=  \mod{\nu}^2 \\
\mbox{and}\
\mod{u}^2& = & {1 \over 2 \beta \cos \gamma} (\Log \varphi)_x.
\end{eqnarray}
If $\gamma = \pi/2$, the Kundu \cite{Kundu1984,Kundu1987} gauge transformation
$u = \nu e^{i\beta \varphi}$ transforms the more general 
higher order nonlinear Schr\"odinger equation (HNLS)
\index{higher order NLS equation (HNLS)}
\begin{equation}
i u_t + u_{xx}  + \delta {\mod{u}}^2 u + (\beta^2 {\mod{u}}^4 +2 i \beta
{\mod{u}}^2)_x)u=0,\ 
(\beta, \delta) \in {\cal R}, 
\end{equation}
into the nonlinear Schr\"odinger equation (NLS)
  \index{nonlinear Schr\"odinger (NLS) equation}
\begin{equation}
i \nu_t + \nu_{xx}  + \delta {\mod{\nu}}^2 \nu = 0. 
\end{equation}
% This transformation is invertible (Calogero and de Lillo)

The natural question is then~:
where do these miraculous transformations from $u$ to another field
come from?
This will be answered in section 
\ref{sectionSingularPartTransformation}. % chapitre.section.soussection MODIF
% ce qui n'est pas le cas par d\'efaut                                  MODIF
% ---------------------------------------------------------------------------
\section{Reduction to ODEs with the Painlev\'e property}
\indent

%In the field of NLPDEs we are interested in, 
Ablowitz, Ramani and Segur \cite{AblowitzSegur1977,ARS1978,ARS1980}, 
and McLeod and Olver \cite{McLeodOlver} 
conjectured a link between integrable NLPDEs and the \Plv\ ODEs 
\cite{PaiCRAS1906}~: 
for the integrable NLPDEs specially studied in these lectures,
all known reductions to ODEs are single\-valued algebraic transforms of
the Weierstrass or \Plv\ equations. 
%  or linearisable
Some of them are listed in Table \ref{TableReductions}.

%\clearpage
%\vfill\eject

\tabcolsep=1.5 truemm                                                % MODIF
\index{KdV equation}
\index{MKdV equation}
\index{nonlinear Schr\"odinger (NLS) equation}
\index{Tzitz\'eica equation}
\begin{table}[h]       % Incoh\'erence entre \label et \ref     MODIF
\caption{Some reductions of a PDE to an ODE and their solutions.
The PDEs $E(u,x,t)=0$ (KdV, MKdV, sG, Bq, NLS, Tzi) are respectively defined 
by the equations (\ref{eqKdVcons}), (\ref{eqMKdVcons}), (\ref{eqsG}),
(\ref{eqBoussinesqAllConventions}), (\ref{eqNLS}), (\ref{eqTzi}). 
The reduction to an ODE for $U(\xi)$ is defined by the two expressions of $u$
in terms of $(U,x,t)$ and of $\xi$ in terms of $(x,t)$. 
The letter $K$, with or without subscript, denotes an arbitrary constant. 
Last column indicates the elementary function ($\wp$, (P1)--(P6)) 
whose general solution of the ODE is a singlevalued algebraic transform.
}\label{TableReductions}
\vspace{0.4cm}
\begin{tabular}{| c | c | c | c | c |}
\hline PDE &
$u$ &
$\xi$ & ODE &
$\wp$, (Pn)
\\
\hline KdV &
$U$ &
$x-ct$ & 
$ \displaystyle \vstrut U'^2 + 2 U^3 - c U^2
  \atop
  \displaystyle \strut + 2 K_1 U + 2 K_2=0$ &
$\wp$
\\
\hline KdV &
$U - \lambda t$ &
$x + 3 \lambda t^2$ &
$ \displaystyle \vstrut U'' + 3 U^2
  \atop
  \displaystyle \strut  - \lambda \xi + K=0$ & (P1)
\\
\hline MKdV &
$U$ &
$x - c t$ &
$ \displaystyle \vstrut U'^2 - U^4 - c U^2 
  \atop
  \displaystyle \strut + K_1 U + K_2=0$ &
$ \wp$
\\
\hline MKdV &
$(3 t)^{-1/3} U$ &
$x (3 t)^{-1/3}$ &
$ \displaystyle \vstrut U'' - 2 U^3 
  \atop
  \displaystyle \strut - \xi U + K = 0$ & (P2)
\\
\hline sG &
$- i \Log U$ &
$x-ct$ &
$ \displaystyle \vstrut c U'^2  + U^3 
  \atop
  \displaystyle \strut + K U^2 + U=0$ &
$\wp$
\\
\hline sG &
$- i \Log U$ &
$x t$ &
$ \displaystyle \vstrut U'' - U'^2 / U + U' / \xi 
  \atop
  \displaystyle \strut + (1 - U^2) / (2 \xi)=0$ & \hbox{(P3)} % particular
\\
\hline Bq &
$U$ &
$x - c t$ &
$ {\displaystyle \vstrut (U'' / 3) + U^2 
  \atop
  \displaystyle  + c^2 U + K_1 \xi}
  \atop
  \displaystyle  \strut + K_2=0$ &
$\wp$, (P1)
\\
\hline Bq &
$2 (U' +\xi- t^2)$ &
$ {\displaystyle x - t^2
  \atop
  \displaystyle + K_1} $ &
$ \displaystyle \vstrut U''^2 + 4 U'^3 
  \atop
  {\displaystyle +12 (\xi U'-U)U'  
  \atop
  \displaystyle \strut + K_2 U' + K_3=0}$  &
$ \displaystyle \hbox{(P2)}
  \atop
  \displaystyle \hbox{\cite{Quispel,ChazyThese}}$
\\
\hline Bq &
$(U'-\xi^2/2)/ t$ &
$x t^{-1/2}$ &
$ {\displaystyle \vstrut U''^2/2 + U'^3 
  \atop
  \displaystyle - (9/8)(U- \xi U')^2}
  \atop
  {\displaystyle +K_1(U- \xi U') 
  \atop
  \displaystyle \strut +K_2 U' +K_3=0}$  &
% M.II (9.25) page 94, puis (16.13) page 104, non int\'egr\'ee par Bureau
$ \displaystyle \hbox{(P4)}
                \hbox{\cite{Quispel}}
  \atop
  \displaystyle \hbox{\cite{BureauMII,ChazyThese}}$
\\
\hline NLS & (\ref{eqNLSReduc1}) &
$ x - c t $ & (\ref{eqNLS45}) &
$\wp$
\\
\hline NLS &
$e^{i(x t - 4 t^3 / 3)} U$ &
$x - t^2$ &
$ \displaystyle \vstrut U'' + 2 \varepsilon U^3 
  \atop
  \displaystyle \strut - 2 \xi U=0$ &
$ \displaystyle \hbox{(P2)}
  \atop
  \displaystyle \hbox{\cite{Tajiri1983}}$
\\
\hline NLS &
$t^{-1/2} \sqrt{U'} e^{i \varphi}$ &
$x t^{-1/2}$ &
$ \displaystyle \vstrut 4 U''^2 + 4 \varepsilon U'^3 + K U'
  \atop
  \displaystyle \strut +(\xi U' - U)^2 / 4 = 0 $ &
$ \displaystyle \hbox{(P4)}
                \hbox{\cite{BoitiPempinelli1980}}
  \atop
% Chazy Th\`ese : (3) p 320, (B-III) p 340
  \displaystyle \hbox{\cite{ChazyThese,BureauMIII}}$
\\
\hline Tzi  &
$ \Log U$  &
$x-ct$  &
$
 \displaystyle \vstrut -c U'^2  + 2 a U^3 
  \atop
 \displaystyle \strut +K U^2 - a_0 = 0 
$  &
$\wp$
\\
\hline  Tzi  &
$ \Log U$  &
$x t$  &
$
 \displaystyle (\xi U'/U)' + a U
  \atop
 \displaystyle + a_0 U^{-2}=0
$ 
  %\atop
  %\displaystyle \strut + (1 - U^2) / (2 \xi)=0$  
& 
$ \displaystyle \hbox{(P3)} % particular
  \atop
  \displaystyle \hbox{\cite{CMOR1986}}$
\\
\hline
\end{tabular}
\end{table} 

% ---------------------------------------------------------------------------
\section{Construction of solitary wave solutions}
\index{solitary wave} 
\indent

In the integrable case, 
the solitary waves
$\sech$ and $\sech^2$ are degenerate elliptic functions,
obtained by imposing % some 
boundary conditions
to the general solution of the ODE defining the travelling wave reduction.
% of the PDE.
In the partially or nonintegrable case, 
the general solution of the reduction may not exist. 
One then looks for particular solutions, 
taking advantage of the singularity structure of the ODE 
by the method of subequations in chapter \ref{chap4}. 
\index{solitary wave} 
\index{elliptic function}

{\it Example1.} {\bf KdV}
\index{KdV equation}

The reduction $\ u(x,t) = U(\xi),\ \xi=x-ct,$ of Eq.~(\ref{eqKdVcons}) 
yields the ODE
\begin{equation}
\label{eqKdVReduc1}
 (-c U + U'' + 3 U^2 )' = 0
\end{equation}
After two integrations, this equation becomes
\begin{equation}
\label{eqKdVredxi}
-c U^2 / 2 + U^3  + U'^2 / 2 + K_1 U + K_2=0,
\end{equation}
which identifies to the Weierstrass elliptic equation
\index{elliptic function}
\begin{eqnarray}
\label{eqWEIE}
{\hskip -3.5 truemm}
 \wp'^2 
{\hskip -3 truemm}
& = &
{\hskip -3 truemm}
4 \wp^3 - g_2 \wp - g_3,\ (g_2,g_3) \hbox{ real constants},
\\
{\hskip -3.5 truemm}
 u
{\hskip -3 truemm}
& = &
{\hskip -3 truemm}
c/6 -2 \wp(x-ct-x_0, c^2/12 - K_1, K_2/2 + K_1 c/12 -(c/6)^3).
\end{eqnarray}
The solitary wave\index{solitary wave} (\ref{eqKdVsw})
is found by imposing the boundary conditions 
$U(\xi) \to 0$,
$U'(\xi) \to 0,\ U''(\xi) \to 0$,
when $\mod{\xi} \to \infty$.
Note that, for $K_1=K_2=0$, 
equation (\ref{eqKdVredxi}) is a degenerate elliptic equation and 
\index{elliptic function}
\begin{equation}
\wp(x-ct-x_0,c^2 / 12,-(c/6)^3) 
= - (c/4) \sech ^2 \left(\sqrt{c}(x-ct-x_0)/2 \right) +c/12.
\end{equation}

%{\it Example sG}.\par
%\noindent 
%The change of variable $ v= e^{\pm i u} $ transforms equation (\ref{eqsG})
%into  the polynomial expression~:
%\begin{equation}
%\label{eqsGP}
% v v_{xt} - v_x v_t - {1\over 2} v^3 + {1\over 2} v = 0
%\end {equation}
%The reduction $(v,x,t) \to (V,\xi) $ yields, after one integration, the ODE~:
%\begin{equation}   
% cV'^2 + V^3 + K V^2 + V = 0,\ K\ \hbox{arbitrary}
%\end{equation}
%with general solution 
%\begin{eqnarray}
% V  &=& {K\over 3 c} - 4 c\ \wp ( x - ct + x_0, g_2, g_3 ) \\
%g_2&=&{1\over 4 c^2}\left({ K^2 \over 3 c^2} + {2 K^2 \over 3 c} + 1\right),\ 
%g_3= -{K\over 3 c} \left({ K^2 \over 9 c^2} + { K^2 \over 3 c} + 1 \right)    
%\end{eqnarray}
%

{\it Example2.} The generalised {\bf Tzitz\'eica} equation
\index{Tzitz\'eica equation}
\begin{equation}
\label{eqGTzi}
  u_{xt} + a e^u + a_1 e^{-u} + a_0 e^{-2u} = 0,\ a \ne 0 
\end{equation}
includes Liouville $(a_1 = a_0 = 0)$, sinh-Gordon $(a_1 \ne 0, a_0 =0)$ 
or Tzitz\'eica \cite{Tzitzeica1908,Tzitzeica1910} $ (a_0 \ne 0, a_1 = 0)$
equations. It is polynomial in the variable $ v = e^u $
\begin{equation}
\label{eqGTziP}
v v_{xt} - v_x v_t + a v^3 + a_1 v + a_0 = 0 
\end{equation}
Its reduction  $ (v,x,t) \to (V,\xi= x -ct) $ can be integrated once
\begin{equation}
- c V'^2 + 2 a V^3 - 6 K V^2 - 2 a_1 V - a_0 = 0,\ K\ \hbox{arbitrary} 
\end{equation}
and possesses the general two-parameter solution \cite{ConteMusette1992}
\begin{equation}
a V = K + 2c
\wp(\xi - \xi_0,(3 K^2 + a a_1)/c^2,(4 K^3 + 2 a a_1 K + a^2 a_0)/(4 c^3)).
\end{equation} 
Moreover a linear superposition of two waves with opposite directions
\begin{equation}
 v(x,t) = A f(x-ct) + B g(x+ct) 
\end{equation}
is compatible  with the Tzitz\'eica \index{Tzitz\'eica} equation by assuming
that $f$ and $g$ satisfy the following second order ODE 
with constant coefficients    
\begin{equation}
 f''= A_1 f^2 + B_1,\
 g''= A_2 g^2 + B_2
\end{equation}
A particular solution of (\ref{eqGTzi}) for $ a_1 = 0 $ is then
\cite{MusetteConte1994}~:
\index{Tzitz\'eica equation}
\begin{eqnarray}
a e^u
&=&
\phantom{-}
  2 c\ \wp(x-ct-x_1, g_2, K + a^2 a_0/(8 c^3))
\nonumber 
\\
& &
- 2c\ \wp(x-ct-x_2, g_2, K - a^2 a_0/(8 c^3)),\
\\
& &
c,x_1,x_2,g_2,K \hbox{ arbitrary constants.}
\nonumber
\end{eqnarray}

{\it Example3.} {\bf NLS}

\noindent The reduction $ u(x,t) = \rho(\xi) e^{i[- \Omega t + \varphi(\xi)]}$
of (\ref{eqNLS}) yields the coupled ODE's
\index{nonlinear Schr\"odinger (NLS) equation}
\begin{eqnarray}
\label{eqNLSReduc1}
 & & -c\rho'+2\varphi' \rho'+ \varphi'' \rho=0,
\\
\label{eqNLSReduc2} 
 & & \rho''+(\Omega-(\varphi')^2+c\varphi')\rho+2 \varepsilon\rho^3=0
\end{eqnarray} 
Equation (\ref{eqNLSReduc1}) admits the integrating factor $\rho$
\begin{equation}
\label{eqNLS98}
\varphi'= c+K_1/S,\ S=\rho^2.
\end{equation}
Then Eq.~(\ref{eqNLSReduc2}) admits the integrating factor $\rho'$, hence
\begin{equation}
\label{eqNLS45}
S'^2 = - 4 \varepsilon S^3 - 4 \alpha  S^2 + 8 K_2  S
 - K_1^2,\ \alpha= \Omega + c^2 / 4,
\end{equation}
an elliptic equation for $S$ with the general solution 
\index{elliptic function}
\begin{eqnarray}
S 
&=& 
- \alpha/(3 \varepsilon) -\wp(x-ct-x_0,g_2,g_3) / \varepsilon
\\
g_2 
&=& 
8  \varepsilon (K_2 + \alpha^2 / (6 \varepsilon)),\
g_3 
=
(2 \alpha / 3)^3 + 8 K_2 \alpha \varepsilon / 3 + \varepsilon^2 K_1^2 .
\end{eqnarray}
The one-soliton solution is obtained for the values of $K_1,K_2$ making
the Weierstrass elliptic function degenerate into a trigonometric
function~:
\begin{equation}
\wp(\xi,g_2,g_3) \to a_1 + a_2 \sech^2 k\xi.
\end{equation}
This happens in two cases~:
\begin{enumerate}
\item
$a_1=K_1=K_2=0,\ k^2=-\alpha, \rho^2 = (k^2 / \varepsilon) \sech ^2 k\xi $
\item 
$a_1K_1K_2\ne 0,\ a_1=-(k^2+\alpha)/ (3 \varepsilon),\
\rho^2= a_1 + (k^2 / \varepsilon) \sech^2 k\xi$.
\end{enumerate}
They respectively correspond for equation (\ref{eqNLS}) to the 
three-parameter $(c,k,x_0)$ solution (``bright" soliton) \cite{ZS1971}
\begin{equation}
 \varepsilon > 0 :\ 
u = \varepsilon^{-1/2} k \sech(k(x-ct-x_0))
 e^{i c x /2 + i (k^2 - (c/2)^2) t}
\label{eqBright}
\end{equation}
and the four-parameter $(c,k,K,x_0)$ solution (``dark" soliton) \cite{ZS1973}
\begin{eqnarray}
  \varepsilon < 0:\ 
u 
& = &
(- \varepsilon)^{-1/2}
 \left[(k/2) \tanh(k (x-ct-x_0)/2) -i (K-c/2)\right]
\nonumber
\\
& &
 \times  
e^{i K x - 2 i [k^2/4 + (K-c/2)^2 + K^2/2]t}.
\label{eqDark}
\end{eqnarray}

% ---------------------------------------------------------------------------
\section{Conservation laws}
\indent

\index{conservation law}
{\it Definition}.
Given a PDE $E(u;x,t)=0$, 
a conservation law is a relation
\begin{equation}
T_t + X_x= 0,
\end{equation}
where $T$ and $X$, respectively called {\it density} and {\it flux},
depend on $x,t,u$ and its derivatives.
If the total variation of $X$ in the interval $a\le x\le b$ is zero,
the quantity $\int_a^b T\ \D x $
is a constant of the motion $I$ called {\it conserved quantity}.
``Integrable'' PDEs possess an infinite number of conservation laws
\cite{AKNS,WSK1975,WKI1979,AblowitzSegur1981,DrazinJohnson}.
For example, the first three conservation laws are~:
\begin{description}
\index{KdV equation}

\item (a) for the KdV equation (\ref{eqKdVcons})
\cite{Whitham1965,MGK1968,KMGZ1970}
\begin{eqnarray}
\hspace{-21.5 truemm} 
& &
T_1=u,\ X_1=3u^2 + u_{xx};
\\ 
\hspace{-21.5 truemm} 
& &
T_2=u^2/2,\ X_2=2 u^3+ u u_{xx}  - u_x^2/2;
\\ 
\hspace{-21.5 truemm} 
& &
T_3=2 u^3 - u_x^2,\ 
X_3=9 u^4 + 6 u^2 u_{xx} - 12 u u_x^2- 2 u_x u_{3x} + u_{xx}^2;   
\end{eqnarray}

\item (b) for the MKdV equation (\ref{eqMKdVcons}) \cite{MGK1968}
\index{MKdV equation}
\begin{eqnarray}
\hspace{-20mm}& &
T_1=u,\ X_1=2 u^3 + u_{xx};\\
\hspace{-20mm}& &
T_2=u^2/2,\ X_2=3 u^4/2 + u u_{xx} - u_x^2/2;\\
\hspace{-20mm}& &
T_3=u^4/4 - u_x^2/4,\ 
X_3=u^6 + u^3 u_{xx} - 3 u^2 u_x^2- u_x u_{3x}/2 + u_{xx}^2/4; 
\end{eqnarray}

\item (c) for the sG equation (\ref{eqsG}) 
\cite{Lamb1971,SCM1973,DoddBullough77}
\begin{eqnarray}
\hspace {-25mm}& &
T_1=u_x^2/2,\ X_1=\cos u;\\
\hspace {-25mm}& &
T_2=u_x^4/4 - u_{xx}^2,\ X_2=u_x^2 \cos u;\\
\hspace {-25mm}& &
T_3= 3 u_x^6 - 12 u_x^2 u_{xx}^2 + 16 u_x^3 u_{3x} + 72 u_{3x}^2,\
X_3=(2 u_x^4 - 24 u_{xx}^2)\cos u. 
\end{eqnarray}

\item (d) for the NLS equation (\ref{eqNLS}), 
we reproduce three of the five conservation laws given by 
Zakharov and Shabat \cite{ZS1971,ZS1973} for $\varepsilon = \pm 1$
\index{nonlinear Schr\"odinger (NLS) equation}
\begin{eqnarray}
\hspace{-9truemm}
&\varepsilon = +1, & 
I_1 = \int_{-\infty}^{+\infty}\mod{u}^2\ \D x,\
I_2 = \int_{-\infty}^{+\infty}(\overline{u} u_x - u\overline{u}_x)\ \D x,
\nonumber
\\  
\hspace{-9truemm}
&\phantom{\varepsilon = +1,}& 
I_3 = \int_{-\infty}^{+\infty}(\mod{u_x}^2 -{1\over 2} \mod{u}^4)\ \D x
\\
\hspace{-9truemm}
&\varepsilon = -1, & 
I_1 = \int_{-\infty}^{+\infty}(1 -\mod{u}^2 )\ \D x,\
I_2 =-\int_{-\infty}^{+\infty}(\overline{u} u_x -u\overline{u}_x)\ \D x,
\nonumber
\\                  
\hspace{-9truemm}
&\phantom{(ii)\ \varepsilon = -1}&\ 
I_3 = \int_{-\infty}^{+\infty}( \mod{u}^4 + \mod{u_x}^2 - 1)\ \D x
\end{eqnarray}
(where $\overline{u}$ denotes the complex conjugate of $u$).
\end{description}

For the Tzitz\'eica equation (\ref{eqTzi}),
Dodd and Bullough \cite{DoddBullough77} first obtained two nontrivial 
conservation laws,
then Mikhailov \cite{Mikhailov1981} gave a recursion formula for an infinite 
set of nontrivial polynomial conserved densities.
\index{Tzitz\'eica equation}
  
% ---------------------------------------------------------------------------
\section{B\"acklund transformations}
\indent

\def\BT{B\"acklund transformation}

\subsection{Definition}
\label{sectionBTDefinition}
\indent

A B\"acklund \cite {Backlund1883} transformation (BT)
\index{B\"acklund transformation}
between two given PDEs 
\begin{equation}
 E_1(u;x,t)=0,\ E_2(v;x',t')=0
\end{equation}
is a set of four relations (\cite{DarbouxSurfaces} vol.~III chap.~XII)
\begin{eqnarray}
& & F_j(u,v,u_x,v_{x'},u_t,v_{t'},\dots;x,t,x',t')=0,\ j=1,2 \\
& & x'=X(x,t,u,u_x,u_t,v),\
 t' = T(x,t,u,u_x,u_t,v)
\end{eqnarray}
such that the elimination of $u$ (resp.~$v$) between $(F_1,F_2)$
implies
\begin{displaymath}
E_2(v;x',t')=0  \hbox{ (resp.~}E_1(u;x,t)=0).
\end{displaymath}
In case the two PDEs are the same, the BT is called an auto-BT.

B\"acklund theory originates from the work of Lie and B\"acklund for the
study of surfaces in differential geometry.
The subject was subsequently developed by Goursat \cite{Goursat1925} 
and Clairin \cite{Clairin}.
% references are in {\it {\BT}s and their applications}, Rogers and Shadwick
{\BT}s re-present an extension of Lie contact transformations. 
They were first obtained for second order PDEs in two independent variables,
linear in the highest derivatives (i.e.~a special type of Monge-Amp\`ere
equation).

For more details on BTs,
the reader is advised to consult the book by Rogers and Shadwick
\cite{RogersShadwick}
and the classical book of Goursat \cite{Goursat}.

% ---------------------------------------------------------------------------
\subsection{Examples~: second order PDEs} 
\indent

\subsubsection{Burgers and heat equations}
% Rogers and Shadwick p 38
\indent
Given the two equations

\index{Burgers equation}
\begin{equation}
E_1 \equiv u_t + (u_x + u^2)_x =0,\
E_2 \equiv v_t + v_{xx}=0.
\end{equation}
the two relations defining the BT are~:
\index{B\"acklund transformation}
\begin{equation}
F_1 \equiv v_x - u v =0,\
F_2 \equiv v_t + u^2 v + v u_x = 0.
\end{equation}
Indeed, the elimination of $v$ (resp.~$u$) yields the identities
\begin{eqnarray}
{\hskip -2 truemm}
(F_2/v)_x-(F_1/v)_t
& \equiv &
E_1,\ v\not=0,\
\hbox{ and }
F_2 + F_{1,x} + u F_1 \equiv E_2.
\end{eqnarray}

\subsubsection{Liouville and d'Alembert}
\indent
Given the two equations
\index{Liouville equation}
\index{d'Alembert equation}
\begin{equation}
 E_1 \equiv u_{xt}-e^u=0,\ E_2 \equiv v_{xt}=0,
\label{eqLiouville}
\end{equation}
the two relations
\begin{eqnarray}
\label{LiouvilleBT}
F_1
& \equiv &
  u_x-v_x + \lambda e^{(u+v) / 2}=0
\label{eqBT1Liouville}
\\
F_2
& \equiv &
  u_t+v_t +(2/ \lambda) e^{(u-v) / 2}=0,
\label{eqBT2Liouville}
\end{eqnarray}
where $\lambda$ is an arbitrary real constant called B\"acklund parameter,
define a \BT\ as shown by the elimination of $v$ (resp.~$u$)
\index{B\"acklund transformation}
\begin{eqnarray}
& &
F_{1,t} + F_{2,x} 
- (1 / \lambda) e^{(u-v) / 2} F_1 - (\lambda/2) e^{(u+v) / 2} F_2
\equiv  2 E_1
\\
& &
F_{1,t} - F_{2,x} 
+ (1 / \lambda) e^{(u-v) / 2} F_1 - (\lambda/2) e^{(u+v) / 2} F_2
\equiv - 2 E_2.
\end{eqnarray}

Thus, the general solution of d'Alembert equation
\begin{equation}
 v=f(x)+g(t),\ (f,g) \hbox{ arbitrary functions},
\end{equation}
provides,
by integration of the ODEs (\ref{eqBT1Liouville})--(\ref{eqBT2Liouville}), 
a solution of (\ref{eqLiouville})
\begin{equation}
 e^u=2 \varphi_x \varphi_t / \varphi^2,\
\varphi=(\lambda/2) \int^x e^f \D x +(1/ \lambda) \int^t e^{-g} \D t
\end{equation}
which is % therefore 
the general solution.
Travelling waves are built by the choice
\begin{equation}
 \varphi= \cotanh(\alpha x) - \tanh(\beta t) 
 \Rightarrow 
 e^u=2 \alpha \beta / \cosh^2(\alpha x - \beta t). 
\end{equation}

\subsubsection{Sine-Gordon}
\indent
Given two solutions $u$ and $U$ of the sine-Gordon equation
\index{sine-Gordon equation}
\begin{equation}
 E_1 \equiv u_{xt} - \sin u=0,\ E_2 \equiv U_{xt} - \sin U=0,
\end{equation}
the auto-\BT\ is defined by
\begin{eqnarray}
{\hskip -6truemm}
 F_1 & \equiv &
 (u+U)_x-2 \lambda \sin((u-U)/2)=0 
\label{sineGordonBTx}
\\
{\hskip -6truemm}
 F_2 & \equiv &
 (u-U)_t-(2/ \lambda) \sin((u+U)/2) =0,\
\lambda \hbox{ arbitrary constant},
\label{sineGordonBTt}
\end{eqnarray}
as can easily be checked quite similarly to the Liouville and d'Alembert case,
by elimination of $U$ (resp.$u$) between these two relations
\begin{eqnarray}
{\hskip -11truemm}
& &
F_{1,t} + F_{2,x} + (1/ \lambda) \cos((u+U)/2) F_1 + \lambda \cos((u-U)/2) F_2
\equiv  2 E_1 \\
{\hskip -11truemm}
& & 
F_{1,t} - F_{2,x} - (1/ \lambda) \cos((u+U)/2) F_1 + \lambda \cos((u-U)/2) F_2
\equiv  2 E_2.
\end{eqnarray}
Lamb \cite{Lamb1967} built from (\ref{sineGordonBTx})--(\ref{sineGordonBTt})
infinite families of % particular 
solutions, 
e.g.~the $N-$soliton solution~:
at the first iteration, one starts from the solution $U=0$ 
(``vacuum''),
\index{vacuum}
and the integration of the ODEs
(\ref{sineGordonBTx})--(\ref{sineGordonBTt})
yields 
\begin{equation}
\tan (u/4) = e^{\lambda x + \lambda^{-1} t + \delta },\ 
\delta \hbox{ arbitrary constant},
\end{equation}
i.e.~the one-soliton solution
\begin{equation}
 u_x=2 \lambda      \sech (\lambda x + \lambda^{-1} t + \delta),\
 u_t=2 \lambda^{-1} \sech (\lambda x + \lambda^{-1} t + \delta).
\end{equation}

\subsubsection{Tzitz\'eica}
\indent
For the Tzitz\'eica equation (Tzi)
\index{Tzitz\'eica equation}
\begin{equation}
\label{eqTzi}
   u_{xt} = e^u - e^{-2u}
\end{equation}
there exists a complicated auto-BT \cite{SaSh,BoSaSh},
and another, much simpler one will be published soon \cite{CGM1998}.
We only report here the classical, well established results.

The Lax pair given by Tzitz\'eica
\cite{Tzitzeica1908,Tzitzeica1910} and rediscovered by Mikhailov
\cite{Mikhailov1979,Mikhailov1981} consists in the following matricial system
of linear PDE's~: 
\begin{eqnarray}
{\partial\over \partial x}
\left(
\begin{array}{c} \varphi\\ \partial_x \varphi\\ \partial_t \varphi\end{array}
\right)&=& 
\left(
\begin{array}{ccc}0&1&0\\ 0 & U_x & \lambda e^{-U}\\ e^U & 0 & 0\end{array}
\right) 
\left(
\begin{array}{c} \varphi\\ \partial_x \varphi\\ \partial_t \varphi\end{array}
\right) \\
{\partial\over \partial t}
\left(
\begin{array}{c} \varphi\\ \partial_x \varphi\\ \partial_t \varphi\end{array}
\right)&=& 
\left(
\begin{array}{ccc}0&0&1\\ e^U & 0 & 0\\ 0 & \lambda^{-1} e^{-U} & U_t
\end{array}\right) 
\left(
\begin{array}{c} \varphi\\ \partial_x \varphi\\ \partial_t \varphi\end{array}
\right)
\end{eqnarray}
while the Moutard \cite{Moutard1875} transformation between two solutions
$U,u$ writes \cite{Tzitzeica1910}~:
\begin{equation}
e^u = - e^U + 2 \varphi_x \varphi_t / \varphi^2.
\end{equation}

Since 1973, BTs
have been found for PDEs of order greater than two. 
Different approaches have been used for deriving those transformations~:
\index{B\"acklund transformation} 
\begin{enumerate}

\item
the method of Clairin \cite{Clairin,Lamb1974}, 

\item
the method of differential forms developed by Wahlquist and Estabrook
\cite{WE1973,WE1975,EW1976}),

\item
the method of bilinear transformations of Hirota 
\cite{Hirota1974,Hirota1980,Matsuno},
\index{bilinear method of Hirota}

\item
the method of gauge transformations developed by 
Boiti {\it et al.~}\cite{BoTu1982,BoLaPeTu1983} and
Levi {\it et al.~}\cite{LeRaSy1982,LeRa1982}. 
\end{enumerate}

In the last two methods, the BT results from the elimination of the wave
function between the Lax pair and the DT.
In next sections,
these two main concepts of complete integrability are briefly recalled;
then the principle of the method of gauge transformations is presented.
Lax pairs and DTs are explicitly given for the PDE's of the AKNS scheme
(KdV, MKdV, sine-Gordon, NLS) 
and for some fifth order PDEs, 
respectively in sections \ref{sectionAKNSScheme} and 
\ref{sectionHigherKdVType}.
But, let us first give some definitions.
\index{Darboux transformation (DT)}
\index{AKNS scheme} 
\index{KdV equation}
\index{MKdV equation}
\index{sine-Gordon equation}
\index{nonlinear Schr\"odinger (NLS) equation}

% ---------------------------------------------------------------------------
\section{Darboux transformation  and Lax pair}
\indent

% ***************************************************** Darboux transformation
\subsection{Definitions}
\indent

\subsubsection{Crum-Darboux transformation}
\indent

This transformation is a key in the theory of nonlinear integrable 
evolution equations for building soliton solutions and understanding their
``asymptotically linear'' superposition rules. 
It is based on a result obtained by the French mathematician Gaston Darboux 
in the special case of the Sturm-Liouville equation 
(also called Schr\"odinger equation in quantum mechanics). 
We briefly recall this old theorem \cite{Darboux1882} and its
generalisation due to Crum \cite{Crum1955}. 
\begin{description}

\index{Darboux transformation (DT)}
\item{\bf Theorem 1.} (Darboux) 
The linear Schr\"odinger equation 
\begin{equation}
\label{eqSchrodinger} 
   \psi_{xx} + ( u + \lambda ) \psi = 0
\end{equation}
is invariant under
\begin{eqnarray}
\label{DT1}
 \psi   & \to & \tilde\psi = (\partial_x - {\psi_{0,x}\over \psi_0}) \psi \\
\label{DT2}
    u   & \to & \tilde u = u + 2 (\Log\psi_0)_{xx} 
\end{eqnarray}
% The involution \psi \to 1/ \psi is the limit \psi_0 \to \psi.
where $\psi_0\equiv\psi(x,\lambda_0)$ is an eigenfunction of 
(\ref{eqSchrodinger}) with parameter $\lambda_0$. 
The essential point is that the new potential $\tilde u$ depends only on
$\psi_0$ and not on $\psi$. 
This transformation can then be iterated to obtain

\index{Crum transformation}
\item{\bf Theorem 2.} (Crum)
The function
\begin{equation}
 \tilde\psi = 
{W(\psi_1,\psi_2,\dots,\ \psi_N,\psi) \over W(\psi_1,\psi_2,\dots,\ \psi_N)}
\end{equation}
where $\psi_1,\psi_2,\dots \psi_N $ are eigenfunctions of 
(\ref{eqSchrodinger}) associated with
parameters $\lambda_1,\lambda_2,\dots,\ \lambda_N$ 
and the symbol $W$ represents the Wronskian determinant, 
solves the equation (\ref{eqSchrodinger}) for the potential
\begin{equation} 
\tilde u = u + 2 \left(\Log W(\psi_1,\psi_2,\dots,\ \psi_N)\right)_{xx}.
\end{equation}
\end{description}

% ******************************************************************* Lax pair
\subsubsection{Lax pair} 
\indent

In 1968  Lax \cite{Lax1968} explained in a very transparent way the greater 
part of the result of Gardner {\it et al.~}\cite{GGKM} 
by introducing the following operators
\index{Lax pair}
\index{KdV equation}
\begin{equation}
\label{LaxKdV}
  L = -\partial^2_x - u(x,t),\ A=-4 \partial^3_x - 6u\partial_x - 3 u_x
\end{equation}
such that the KdV equation (\ref{eqKdVcons}) may be represented 
in the following way
\begin{equation}
\label{Laxrepresentation}
\partial_t L = [ A, L]   
\end{equation}
called the Lax representation. 
Equation (\ref{Laxrepresentation}) expresses the compatibility 
between the two partial differential equations of the system
\begin{equation}
\label{Laxpair}
\left. % \left\{
\begin{array}{rcl}
L\psi
\hspace{-3mm} 
&=& 
\hspace{-3mm}
\lambda \psi  
\nonumber
\\
\psi_t
\hspace{-3mm}
&=& 
\hspace{-3mm}
A \psi       
\nonumber
\end{array}
\right\} % \right.
 \iff
\left\{
\begin{array}{l}
\psi_{xx}+( u + \lambda) \psi = 0 \\
\psi_t + (2 u - 4 \lambda)\psi_x -  u_x \psi = 0
\end{array}
\right.
\end{equation}
called Lax pair. This equivalence results from the identity
\begin{equation}
\psi_{xxt} - \psi_{txx} \equiv \hbox{KdV}(u) \psi
\end{equation}
The system (\ref{Laxpair}) is invariant
under the Darboux transformation (\ref{DT1})--(\ref{DT2}) 
with the compatibility condition
\index{Darboux transformation (DT)}
\begin{equation} 
   (\partial_t\tilde L)\tilde\psi = [\tilde A,\tilde L]\tilde \psi
\end{equation}
where $(\tilde L,\tilde A) $ results from the substitution of $u$ by 
$\tilde u$ in $(L,A)$.

%*******************************B\"acklund gauge transformation***************
\subsection{B\"acklund gauge transformation}
\indent

\index{B\"acklund gauge transformation}
\index{B\"acklund transformation}
A general procedure to obtain BT for nonlinear PDE's derived as 
compatibility conditions between a given generalised Lax pair of operators
was simultanuously considered by Boiti {\it et al.~}and Levi {\it et al.~}in 
1982. 
It has provided new results for multidimensional nonlinear PDE's.
Here, we only report the principle of the method. 
Let us consider the Lax pair
\begin{equation}
\label{gLaxpair}
\psi_x = L \psi,\ \psi_t = M \psi
\end{equation}
where $\psi$ is a $N \times N$ matrix as well as $L,M$
which have a preassigned dependence on a matrix ``potential'' $Q(x,t)$
and on a constant parameter $\lambda$. 
The compatibility condition between the two equations of the system 
(\ref{gLaxpair}) implies the following nonlinear equation
\begin{equation}
\label{nlpde}
      L_t - M_x + [ L, M ] = 0.
\end{equation}
To construct the BT for this nonlinear {\pde} one has to consider two 
different systems of type (\ref{gLaxpair}) corresponding to two different
``potentials'', say $ Q(x,t) $ and $\tilde Q(x,t)$~:
\begin{eqnarray}
\label{gLaxpair1}
\psi_x = L(Q(x,t);\lambda) \psi,
&\ \ &
\psi_t = M(Q(x,t);\lambda) \psi \\
\label{gLaxpair2}   
\tilde\psi_x = \tilde L(\tilde Q(x,t);\lambda)\tilde\psi,
&\ \ &
\tilde\psi_t = \tilde M(\tilde Q(x,t);\lambda)\tilde\psi
\end{eqnarray}
One assumes that the following generalised DT holds between the wave functions
$\psi$ and $\tilde\psi$~:
\begin{equation}
\label{gDT}
\tilde\psi = B \psi
\end{equation}
where $B$ is a matrix function of $Q, \tilde Q,x,t$ and $\lambda$. 
The compatibility between (\ref{gDT}) and the system 
(\ref{gLaxpair1})--(\ref{gLaxpair2}) gives the auto-BT
\begin{equation}
B_x = \tilde L B - B L,\ B_t = \tilde M B - B M
\end{equation}
By cross-differentiating these two relations one gets~:
\begin{equation}
(\tilde L_t - \tilde M_x + [\tilde L,\tilde M ]) B - B(L_t - M_x + [L,M]) = 0
\end{equation}                 
which implies that if $ Q(x,t)$ satisfies the nonlinear PDE (\ref{nlpde}) then
$\tilde Q(x,t)$  satisfies the same equation. 
This exactly coincides with the definition of the BT previously given 
in section \ref{sectionBTDefinition}.

Let us also mention the book of Matveev and Salle \cite{MaSa1991}
as a basis reference on Darboux transformation and its development in
soliton theory.    
\index{Darboux transformation (DT)}

% Comment on what a good Lax pair is?

%For a given class of NLPDEs (the Zakharov-Shabat or AKNS scheme, 
% see details below), Lax pairs
%have been related \cite{WSK1975} to B\"acklund transformations  and to the 
%existence of an infinite number of conservation laws.

In the extension of Painlev\'e analysis to NLPDEs \cite{WTC}, 
if a PDE fulfills the necessary conditions of integrability (``\Plv\ test''),
\index{Painlev\'e test for PDEs}
one tries to determine a Lax pair and a Darboux 
transformation relating two solutions of the same PDE
in order to constructively prove the sufficiency of these conditions. 
A method (truncation procedure) leading to such a Lax pair and DT will be 
explained in section \ref{sectionWeissLimitations}.
In this formalism, the link with the notion of ``general solution'' is that 
the knowledge of the BT {\it a priori} allows to build wide classes of 
solutions. % of physical interest. 
In one space dimension the ``good'' Lax pair of a given nonlinear PDE
must depend on the solution of this equation and an arbitrary constant
% parameter 
$\lambda$. 
In the next section we show on examples how to derive the
Lax pair and DT from the associated BT.
In each case, it will be the aim of these lectures to show 
in chapter \ref{chap3} how these two informations can be found
algorithmically by singularity analysis.

% ---------------------------------------------------------------------------
\subsection{Examples~: AKNS scheme}
\index{AKNS scheme}
\label{sectionAKNSScheme}
\indent

% --------------------------------------------------------------- CDT for KdV
\subsubsection{\KdV}
\indent

% Formulae Musette and Conte, J.~Phys.~A 1994 

\index{KdV equation}
Its conservative form is (\ref{eqKdVcons})
and we define the potential form as
\begin{equation}
\label{eqKdVpot}
u=w_x,\ F(w) \equiv w_t + w_{xxx} + 3 w_x^2 =0.
\end{equation}
Given two solutions $w$ and $W$ of (\ref{eqKdVpot}),
the auto-BT is defined by \cite{Lamb1974}
\index{B\"acklund transformation} 
\begin{eqnarray}
(w + W)_x 
& = & 2 \lambda - (w-W)^2/2
\label{eqBTKdV1}
\\
(w + W)_t 
& = & - 2 (w_x^2 + w_x W_x + W^2_x) - (w-W) (w-W)_{xx},
\label{eqBTKdV2}
\end{eqnarray}
where $\lambda$ is the B\"acklund parameter.
After changing variables $w,W$ to % \hfill\break
$W,Y=(w-W)/2$, the gradient of $Y$ is defined by the Riccati equations
\index{Riccati equation}
\begin{eqnarray}
Y_x 
& = & \lambda - U - Y^2,\
U=W_x
\\
Y_t 
& = & \left(U_{x} - (2 U - 4 \lambda) Y\right)_x.
\end{eqnarray}
The transformation
\begin{equation}
\label{eqKdVa}
 Y= \partial_x \Log \psi
\end{equation}
linearises these Riccati equations into one second order ODE
and one first order PDE
\begin{eqnarray}
\label{eqKdVLaxX}
{\hskip -5 truemm}
& & \psi_{xx} + (U - \lambda) \psi=0
\\
{\hskip -5 truemm}
& & \psi_t + (2 U + 4 \lambda) \psi_x - (U_x + G(t)) \psi=0,\
G \hbox{ arbitrary function}.
\label{eqKdVLaxT}
\end{eqnarray}
The Lax pair of KdV is defined by these two linear equations,
which satisfy the compatibility condition
\begin{equation}
\psi_{xxt} - \psi_{txx} = E(U) \psi,
\end{equation}
while the DT for KdV is defined by the $x-$derivative of Eq.~(\ref{eqKdVa})
\begin{equation}
\label{eqDTKdV}
u-U=2 \partial_x^2 \Log \psi.
\end{equation}

% --------------------------------------------------------------- CDT for MKdV
\subsubsection{Modified \KdV}
\indent

% Formulae Musette and Conte, J.~Phys.~A 1994 

Its conservative form is
\index{MKdV equation}
\begin{equation}
\label{eqMKdVcons}
E(u) \equiv u_t + u_{xxx} - 2 a^{-2} (u^3)_x=0
\end{equation}
and we define the potential form as
\begin{equation}
\label{eqMKdVpot}
u=w_x,\ F(w) \equiv w_t + w_{xxx} - 2 a^{-2} w_x^3 =0.
\end{equation}

Given two solutions $w$ and $W$ of (\ref{eqMKdVpot}),
the auto-BT is given by \cite{Lamb1974}
\index{B\"acklund transformation}
\begin{eqnarray}
(w + W)_x 
& = & 
- 2 a \lambda \sinh((w-W)/a)
\label{eqBTMKdV1}
\\
(w + W)_t 
& = & 
   8 \lambda^2 W_x - 4 \lambda W_{xx} \cosh ((w-W)/a)
\nonumber \\ & &
+4 a (2 \lambda^3 - \lambda W_x^2/a^2)\sinh ((w-W)/a),
\label{eqBTMKdV2}
\end{eqnarray}
where $\lambda$ is the B\"acklund parameter.
The change of variables 
\begin{equation}
(w,W) \to (W,Y=e^{(w-W)/a})
\end{equation}
maps these equations into the two Riccati equations for $Y$,
\index{Riccati equation}
\begin{eqnarray}
{\hskip -7 truemm}
Y_x
& = &
- 2 (U/a) Y + \lambda (1 - Y^2) ,\
U=W_x
\label{eqMKdVYx}
\\
{\hskip -7 truemm}
Y_t
& = &
2 A_1 Y + B_1 (1+Y^2) + C_1 (1-Y^2)
\label{eqMKdVYt}
\\
{\hskip -7 truemm}
\phantom{Y_t}
& = &
(- 4 \lambda U/a + (2 (U/a)^2 - 4 \lambda^2 + 2 (U_x/a)) Y)_x
\\
{\hskip -7 truemm}
A_1
& = &
{U_{xx} \over a} - 2 {U^3 \over a^3} + 4 \lambda^2 {U \over a},\
B_1= - 2 \lambda {U_x \over a},\
C_1= 2 \lambda {U^2 \over a^2} - 4 \lambda^3.
\end{eqnarray}
The compatibility condition of this ``Riccati pseudopotential'' $Y$ is
\begin{equation}
 Y_{xt} - Y_{tx} = - (2/a) E(U) Y.
\end{equation}

The Lax pair is obtained by linearising these two Riccati equations
by the transformation 
\begin{equation}
\label{eqMKdV7}
Y=\psi_1/ \psi_2
\end{equation}
\begin{eqnarray}
{\hskip -1 truemm}
& &
\pmatrix{\psi_1 \cr \psi_2 \cr}_x
=
\pmatrix{-U/a & \lambda \cr \lambda & U/a \cr}
\pmatrix{\psi_1 \cr \psi_2 \cr},\
\\
& &
\pmatrix{\psi_1 \cr \psi_2 \cr}_t
=
\pmatrix{A_1 & B_1+C_1 \cr B_1-C_1 & - A_1 \cr}
\pmatrix{\psi_1 \cr \psi_2 \cr}
\end{eqnarray}
while the Darboux transformation is defined by 
\index{Darboux transformation (DT)}
\begin{equation}
\label{eqDTMKdV}
u-U = a \partial_x \Log Y,
\end{equation}
which, by elimination of $Y_x$ with (\ref{eqMKdVYx}), is identical to
\begin{equation}
\label{eqMKdVDT2}
u+U=a \lambda (Y^{-1} - Y).
\end{equation}
The homographic transformation
with $\alpha = U/a $
\begin{equation}
\label{eqY}
Y= \lambda \chi /(1 +\alpha \chi),
\end{equation}
maps the Riccati system (\ref{eqMKdVYx})--(\ref{eqMKdVYt}) 
into the simpler form~:
\begin{eqnarray}
& &
{\hskip -7 truemm}
\chi_x = 1 + (S/2) \chi^2 
\label{eqchix} % ? double emploi avec \label{eqChiX}
\\
& &
{\hskip -7 truemm}
\chi_t = -C + C_x \chi -(1/2) ( C S + C_{xx} ) \chi^2
\label{eqchit}
\\
& & 
{\hskip -7 truemm}
S = 2 \left( {U_x\over a} - \left({U\over a}\right)^2 -   \lambda^2 \right),\ 
C = 2 \left( {U_x\over a} - \left({U\over a}\right)^2 + 2 \lambda^2 \right).
\label{eqSCMKdV}
\end{eqnarray}
We shall see that the relation between the two functions $S$ and $C$~:
\begin{equation}
\label{SMKdV}
 S - C + 6 \lambda^2 = 0
\end{equation}
corresponds to the singular manifold (SM) equation of the KdV equation 
\cite{WTC} 
\index{KdV equation}
\index{singular manifold equation}
and can be found algorithmically \cite{Pickering1996} 
when one performs the Painlev\'e analysis of the MKdV equation.
In the variable 
\begin{equation}
f = a (Y-1)/(Y+1),
\end{equation}
the system (\ref{eqMKdVYx})--(\ref{eqMKdVYt}) 
and the DT (\ref{eqDTMKdV}) or (\ref{eqMKdVDT2}) become
\cite{Wadati1974,WSK1975} 
\begin{eqnarray} 
u-U  & = &  2 a^2 f_x /(a^2 - f^2), % ? non v\'erifi\'e
\label{eqMKdVDT4}
\\
u+U  & = &  - 4 a^2 \lambda f /(a^2 - f^2), % ? non v\'erifi\'e
\label{eqMKdVDT5}
\\
a f_x & = & - (U/a) (a^2 - f^2) - 2 \lambda a f
\\ 
a f_t & = & A_1 (a^2 - f^2) + B_1 (a^2 + f^2) - 2 C_1 a f.
\end{eqnarray}

%{\it Remark}.
%In the above two examples (KdV, MKdV),
%we have used two different transformations to linearise a Riccati system~:
%(\ref{eqKdVa})  which involves one function $\psi$,
%(\ref{eqMKdV7}) which involves two functions $\psi_1, \psi_2$.
%The choice of the linearising transformation is not made at random,
%it is dictated by the singularity structure of the PDE,
%as we will see in sections \ref{sectionTruncationKdV} and 
%\ref{sectionTruncationMKdV}.

% --------------------------------------------------------------- CDT for sG
\subsubsection{Sine-Gordon}
\indent

% Formulae Musette and Conte, J.~Phys.~A 1994 

\index{sine-Gordon equation}
\begin{equation}
\label{eqsG}
E(u) \equiv u_{xt} - \sin u=0.
\end{equation}
Given two solutions $u$ and $U$ of (\ref{eqsG}),
the auto-BT
is given by \cite{Lamb1967}
\index{B\"acklund transformation} 
\begin{eqnarray}
(u + U)_x 
& = & - 4 \lambda \sin((u-U)/2)
\label{eqBTsG1}
\\
(u - U)_t 
& = & - \lambda^{-1} \sin((u+U)/2),
\label{eqBTsG2}
\end{eqnarray}
where $\lambda$ is the B\"acklund parameter.
The change of variables 
\begin{equation}
(u,U) \to (U,Y=e^{-i(u-U)/2})
\end{equation}
maps these equations into the two Riccati equations for $Y$
\index{Riccati equation}
\begin{eqnarray}
\label{eqsGYx}
Y_x
& = & 
i U_x Y + \lambda (1 - Y^2) 
\\
\label{eqsGYt}
Y_t
& = & 
((1-Y^2) \cos U + i (1+Y^2) \sin U)/(4 \lambda).
\end{eqnarray}
The compatibility condition of the Riccati pseudopotential $Y$ is
\begin{equation}
 Y_{xt}-Y_{tx}=i E(U) Y.
\end{equation}

The Lax pair is obtained by linearising the Riccati system
% by the transformation 
\begin{equation}
Y={\psi_1 \over \psi_2}
\end{equation}
\begin{equation}
\pmatrix{\psi_1 \cr \psi_2 \cr}_x
=
\pmatrix{i U_x/2 & \lambda \cr \lambda & - i U_x/2 \cr}
\pmatrix{\psi_1 \cr \psi_2 \cr}
{\hskip -1.0 truemm}
,
\pmatrix{\psi_1 \cr \psi_2 \cr}_t
=
{1 \over 4 \lambda} \pmatrix
 {0 & e^{i U} \cr e^{-i U} & 0 \cr}
\pmatrix{\psi_1 \cr \psi_2 \cr}
\nonumber
\end{equation}
while the Darboux transformation for sG is defined by 
\index{Darboux transformation (DT)}
\begin{eqnarray}
\label{eqDTsG}
u-U
& = &  2 i \Log Y,
\\
\label{eqsGDT2}
(u+U)_x
& = &  2 i \lambda (Y^{-1} - Y). % ? non v\'erifi\'e
\end{eqnarray}
The homographic transformation (\ref{eqY})
with $\alpha = -i U_x/2 $
maps the Riccati system (\ref{eqsGYx})--(\ref{eqsGYt})
into the simpler form (\ref{eqchix})--(\ref{eqchit}), with
\begin{eqnarray}
& &
S=-i U_{xx} + U_x^2/2 - 2 \lambda^2,\ C=-e^{iU}/(4 \lambda^2).
\label{eqSCsineG}
\end{eqnarray}
The relation between $S$ and $C$ 
\index{singular manifold equation}
\begin{equation}
\label{SMsineG}
  S + C_{xx}/C - (1/2) (C_x/C)^2 + 2 \lambda^2 = 0
\end{equation}
represents the SM equation obtained by Conte \cite{Conte1989}
when performing the invariant \Plv\ analysis of the sine-Gordon equation.

%------------------------------------------------------------CDT for NLS
\subsubsection{Nonlinear Schr\"odinger}
\indent

\index{nonlinear Schr\"odinger (NLS) equation}
\begin{equation}
\label{eqNLS} E(u) \equiv i u_t + u_{xx} + 2 \varepsilon \mod{u}^2 u =0,\ 
\varepsilon=\pm 1.
\end{equation} Given two solutions $u$ and $u'$ of (\ref{eqNLS}), the
auto-BT can be written as 
\cite{Chen1974,Lamb1974,KonnoWadati1975,LeviRagniscoSym1984}
\index{B\"acklund transformation}
\begin{eqnarray}
\label{eqNLSBTx} 
(u+U)_x & = & (u-U)
\sqrt{4 \lambda^2 - \varepsilon \mod{u+U}^2}
\\
\label{eqNLSBTt} 
(u+U)_t & = & i (u-U)_x
\sqrt{4 \lambda^2 - \varepsilon \mod{u+U}^2}
\nonumber
\\ 
& & + i \varepsilon  (u+U) (\mod{u+U}^2 + \mod{u-U}^2)/2.
\end{eqnarray}

The extension to NLS of the transformation (\ref{eqMKdVDT5}) is
\index{Darboux transformation (DT)}
\begin{equation}
\label{eqNLSDT1} 
u+U=- 4 \lambda f /(1 + \varepsilon \mod{f}^2).
\end{equation} 
Therefore, the change of variables $(u,U) \to (U,f)$
transforms (\ref{eqNLSBTx}) into
\begin{equation}
 - f_x + \varepsilon f^2 {\overline{f}}_x  =
 (1 - \varepsilon \mod{f}^2) (U (1 +\varepsilon \mod{f}^2) + 2 \lambda f).
\end{equation} 
The elimination of $\overline{f}$ between this equation and its c.c.,
assuming $1 - \mod{f}^4 \not=0$, provides
\begin{equation}
\label{eqNLSfx} 
f_x = - 2 \lambda f - U - \varepsilon {\overline{U}} f^2.
\end{equation} 
while the $t-$part is
\begin{equation}
\label{eqNLSft} 
f_t  = (\lambda U + U_x) + (\varepsilon U \overline{U} + \lambda^2) f
     + (\lambda \overline{U} - \overline{U}_x) f^2
\end{equation} 
with the identity
\begin{equation}
f_{xt} - f_{tx}  = E + \varepsilon \overline{E} f^2.
\end{equation}
Equations (\ref{eqNLSBTx}) and (\ref{eqNLSDT1}) imply
\begin{equation}
\label{eqNLSDT2} 
u-U=2 (f_x - f^2 {\overline{f}}_x) / (1 - \mod{f}^4).
\end{equation}

In all the above examples (KdV, MKdV, sG, NLS), the DT is defined with one
(for KdV) or two (for the others) entire functions $\psi$.
This distinction is the only relevant feature needed to obtain in an 
algorithmic way the Lax pair by methods linked to the singularity
structure of these equations.

% ---------------------------------------------------------------------------
\subsection{Higher order KdV-type equations}
\label{sectionHigherKdVType}
\indent
 
Among the fifth order nonlinear evolution equations 
\begin{equation}
u_t + (u_{xxxx} + (8\alpha - 2\beta) u u_{xx} - 2 (\alpha + \beta) u_x^2 
-(20/3) \alpha \beta u^3)_x = 0  
\end{equation}
only three cases are integrable~:
\index{Kaup-Kupershmidt equation}
\index{Sawada-Kotera equation}
\begin{eqnarray}
\label{eqSK} 
& &
\hspace{-15mm}
\beta / \alpha= -1\ : u_t 
 + 
\left( u_{xxxx} + 10\alpha u u_{xx} + 20\alpha^2 u^3/3 \right)_x = 0 \\
\label{eqKdV5}
& &
\hspace{-15mm}
\beta / \alpha= -6\ : u_t 
+
\left(u_{xxxx} +20\alpha u u_{xx} +10\alpha u_x^2 +40\alpha^2 u^3 \right)_x=0
\\
\label{eqKK}
& &
\hspace{-15mm}
\beta/ \alpha= -16\ : u_t 
+
\left(u_{xxxx} +40\alpha u u_{xx} +30\alpha u_x^2 +320 \alpha^2 u^3/3\right)_x
= 0
\end{eqnarray}
respectively named
Sawada-Kotera (SK) or Caudrey-Dodd-Gibbon \cite{SaKo1974,CaDoGi1976},
Lax's 5th order KdV (KdV5) \cite{Lax1968} and 
Kaup-Kupershmidt (KK) \cite{Kaup1980}.

Their respective Lax representation (\ref{Laxrepresentation})
is \cite{Lax1968,FoGi1980,Fordy1991} 
\begin{eqnarray}
& &
\hspace{-14mm}
\hbox{(SK) }
\alpha = 3,\
L =\partial_x^3 + 6 u \partial_x,\ 
\label{eqLax3SK} 
\\  
& & 
\hspace{-14mm}
A = 9\partial_x^5  + 90
u\partial_x^3  + 90 u_x\partial_x^2  +  (60 u_{xx} + 180 u^2)\partial_x;
\nonumber
\\                            
& &
\hspace{-14mm}
 \hbox{(KdV5) }
\alpha = 1/2,\
L =\partial_x^2 + u,
\nonumber 
\\ 
& & 
\hspace{-14mm}
A = 16\partial_x^5 + 40 u \partial_x^3 + 60 u_x \partial_x^2
 + (50 u_{xx} + 30 u^2) \partial_x + 15 u_{xxx} + 30 u u_x;
\nonumber 
\\                        
& &
\hspace{-14mm}
 \hbox{(KK) }
\alpha = 3/4,\
L =\partial_x^3 + 6 u \partial_x + 3 u_x,
\label{eqLax3KK}
\\ 
& & 
\hspace{-14mm}
A = 3 (3 \partial_x^5 + 30 u \partial_x^3 + 45 u_x \partial_x^2
 + (35 u_{xx} + 60 u^2)\partial_x + 10 u_{xxx} + 30 u u_x).
\nonumber 
\end{eqnarray}
\index{Lax pair of second order}
\index{Lax pair of third order}
We discard the equation (\ref{eqKdV5}) for it has the same second order
scattering problem as the KdV equation,
and we restrict to the two equations (\ref{eqSK})--(\ref{eqKK})
possessing two different third order scattering problems 
$ L \psi =\lambda\psi $.

The SK equation possesses the Darboux transformation
\cite{SaKa1977}
\index{Darboux transformation (DT)}
\index{Sawada-Kotera equation}
\begin{equation}
u = U +  \partial_x^2 \Log\psi
\end{equation}
while for the KK equation this transformation writes 
\cite{LeRa1988}
\index{Kaup-Kupershmidt equation}
\begin{equation}
u = U + (1/2) \partial_x^2 \Log\varphi,\ 
\varphi= \psi\psi_{xx}-(1/2) \psi_x^2 + 3 U \psi^2.
\end{equation}
In the notation $w_x = u,  W_x= U$, the $x-$part of the BT
for SK \cite{DoGi1977,SaKa1977} is
\index{B\"acklund transformation} 
\begin{equation}
\label{eqBTSK}
(w-W)_{xx} + 3 (v-W) (w+W)_x + (w-W)^3 = \lambda,
\end{equation}
while for the KK equation it writes
\begin{equation}
\label{eqBTKK}
(w-W)_{xx} + 3 (w-W) (w+W)_x - (3/4)(w-W)_x^2/(w-W) + (w-W)^3 = \lambda.
\end{equation}
This last expression was obtained for the first time by Rogers and Carillo 
\cite{RoCa1987}
in the particular case $\lambda = 0$.
% ? reference to the BT in bilinear form

% ==========================================================================
\chapter{Painlev\'e analysis for PDEs}
\label{chap3}
\indent

The WTC extension \cite{WTC} of \Plv\ analysis to \pdes\ consists of two parts
\begin{enumerate}
\item 
generation of {\it necessary conditions} (\Plv\ test) for the absence of 
movable critical singularities in the ``general solution'',

\item 
explicit proof of {\it sufficiency} by finding the transformation which
linearises the PDE or yields an auto-BT
or a BT to another PDE with the PP. 
\index{B\"acklund transformation}
\end{enumerate}
The methods relative to both parts are different.

\index{Painlev\'e test for PDEs}
In the first part,
for every noncharacteristic manifold $(\varphi(x,t)=0,\varphi_x\ne 0)$,
one tests the existence of 
all possible local representations of the ``general solution'' by a
Laurent series in the neighbourhood of $\varphi = 0$.
This test may
\begin{itemize}
\item{}
pass whatever be $\varphi$; the PDE may then have the PP,

\item{}
fail whatever be $\varphi$; this is typical of chaotic PDEs,

\item{}
pass with some constraints on $\varphi$;
then there exists particular Laurent series
and the PDE is called ``partially integrable''.

\end{itemize}

\index{Weiss truncation}
In the second part, the Weiss {\it truncation procedure} \cite{Weiss1983},
using only the singular part of the Laurent series,
may yield constructive results like
\begin{itemize}

\item{ } 
the linearising transformation or the BT,
% (i.e.~the DT linking two solutions of the PDE and its Lax pair)
in case the PDE passes the \Plv\ test for every $\varphi$, 
\index{B\"acklund transformation} 

\item{ }
particular solutions, 
necessarily compatible with the singularity structure of the PDE,
in case the \Plv\ test is conditionally or not satisfied 
(see chapter \ref{chap4}).
\end{itemize}

% ---------------------------------------------------------------------------
\section{Necessary conditions (Painlev\'e test)}
\indent

\index{Painlev\'e test for PDEs}
Contrary to the case of ODEs,
the singularities in the complex domain of $(x,t)$ are not isolated.
Given a PDE $E(u,x,t)=0$ of order $N$ polynomial in $u$ and its partial 
derivatives (maybe after a preliminary change of variables),
we consider the associated equation $ \varphi(x,t)=0$ of the movable SM 
\index{singular manifold equation}
and an expansion % and a formal expansion 
of $u$ and $E$ as a Laurent series in $\chi$ 
in the neighborhood of $\varphi=0$.
We distinguish between $\varphi$ and the expansion variable $\chi$
and only require $\chi$ to vanish as $\varphi$
\begin{equation}
u(x,t)  =\sum_{j=0}^{+ \infty} u_j(x,t) \chi^{j+p},\
E(u,x,t)=\sum_{j=0}^{+ \infty} E_j(x,t) \chi^{j+q},
\end{equation}
where $(p,q)$ are two negative integers with $q \le p-1$,
and $(u_j, E_j)$ the Laurent series coefficients.
The result of the \Plv\ test (necessary conditions)
is independent of the explicit expression for 
$\chi$ but some particular choices are better than others 
during the second part (sufficient conditions) when one looks for the Lax
pair or tries to linearise the equation.

The main choices (gauges) for the expansion variable $\chi$ are
\begin{itemize}

\item{i)}
\underbar{WTC gauge} \cite{WTC}
$\chi=\varphi$,
hence coefficients $(u_j,E_j)$ rational in the derivatives 
$D\varphi$ of $\varphi$

\item{ii)}
\underbar{dimensionless WTC gauge} 
$\chi=\varphi/ \varphi_x$,
hence coefficients $(u_j,E_j)$ rational in the derivatives 
$D\varphi$ of $\varphi$ of homogeneity degree zero,

\item{iii)}
\underbar{Kruskal gauge} \cite{JKM}
$\chi=x - f(t),\ f $ arbitrary,
hence coefficients $(u_j,E_j)$ independent of $x$ and
rational in the derivatives of $f$.
This is the simplest choice for the test,
but it cannot be used to obtain the Lax pair or particular solutions.

\item{iv)}
\underbar{Conte gauge} \cite{Conte1989}
$\chi=
\varphi / (\varphi_x- \varphi_{xx}  \varphi/ (2 \varphi_x))
\sim_{\varphi \to 0} \varphi / \varphi_x$,
% MODIF mettre \varphi \to 0 exactement sous \sim
hence coefficients $(u_j,E_j)$ rational in the derivatives
of $\varphi$ invariant under the group of homographic transformations
\begin{displaymath}
\varphi \to (a \varphi + b)/(c \varphi + d),\
(a,b,c,d)\ \hbox{arbitrary complex constants}.
\end{displaymath}
\end{itemize}

In this last case, the Riccati system satisfied by $\chi$ is
\index{Riccati equation}
\begin{eqnarray}
\chi_x 
& = &
 1 + (S/2) \chi^2
\label{eqChiX}
\\
\chi_t 
& = &
 - C + C_x \chi  - (1/2) (C S + C_{xx}) \chi^2
\label{eqChiT}
\\
2 ((\chi^{-1}_t)_x - (\chi^{-1}_x)_t)
& = &
 S_t + C_{xxx} + 2 C_x S + C S_x = 0
\label{eqCrossXT}
\end{eqnarray}
with
\begin{equation}
S = \lbrace \varphi ;x\rbrace
 =  
(\varphi_{xx}/ \varphi_x)_x 
- (1/2) (\varphi_{xx} / \varphi_x)^2,\
C =
- \varphi_t / \varphi_x.
%\label
\end{equation}

The transformation $\chi=\psi / \psi_x$ linearises this Riccati system
into
\index{Lax pair of second order}
\begin{eqnarray}
\label{eqLinearOrder2X}
& &
\psi_{xx} + (S/2) \psi=0
\\
& &
\psi_t + C \psi_x - (C_x/2 + g(t)) \psi=0,\
g \hbox{ arbitrary function}.
\label{eqLinearOrder2T}
\end{eqnarray}
This choice of gauge is equivalent to the % formal 
expansion of
$(u,E)$ as
\begin{equation}
u=\sum_{j=0}^{+ \infty} u_j (\psi/ \psi_x)^{j+p},\
E=\sum_{j=0}^{+ \infty} E_j (\psi/ \psi_x)^{j+q},
\end{equation}
where the function $\psi$ satisfies a second order linear ODE in the $x$ 
variable.

To obtain the couples $(u_0,p)$ one substitutes in the polynomial PDE
\begin{equation}
u \to u_0 \chi^p,\
\chi_x \to 1,\
\chi_t \to -C,\
Du \to u_0 D(\chi^p).
\end{equation}
One then determines the balance between the different terms of this polynomial
expression.
Each different solution $(u_0,p)$ defines a {\it family}.
For every $j\ge 1$ the recurrence relation determining $u_j$ is
\begin{equation}
\forall j \ge 1:\ P(j) u_j=Q_j(\lbrace u_k,Du_k,\ k \in [0,j-1] \rbrace),
\end{equation}
where $P$ is a polynomial of degree at most $N$.

The main requirements of the \Plv\ test are
\index{Painlev\'e test for PDEs}
\begin{itemize}
\item
the zeros of $P$ ({\it Fuchs indices}, also named {\it\Plv\ resonances}) are 
distinct integers,
\item
for every index $i$ and every $\varphi$, 
the compatibility condition $Q_i=0$ holds.
\end{itemize}

% ---------------------------------------------------------------------------
\section{Methods for proving sufficiency}
\indent

One distinguishes two main methods~:
\begin{enumerate}
\item   
{\it the singular part transformation}  which may provide the explicit
transformation linearising the nonlinear PDE. 
If this is not the case, the transformation may
yield an equation in a form more convenient than the original one to search
for explicit solutions,

\item 
{\it the truncation procedure} of Weiss and its extensions for obtaining the
BT and thus proving that the nonlinear PDE possesses the PP. 
\index{B\"acklund transformation} 
\end{enumerate}

% ---------------------------------------------------------------------------
\subsection{Singular part transformation}
\label{sectionSingularPartTransformation}
\indent

The method consists of transforming the PDE for $u$ into an equation for
$\varphi$ by the nonlinear transformation

\begin{equation}
\label{eqSingularPartTransformation}
u= {\cal D} \Log \varphi,
\end{equation}
where ${\cal D}$ is the singular part operator associated with one of the
families defined  in the Painlev\'e test. 
\index{Painlev\'e test for PDEs}

{\it Example} 1 (linearisation).
{\bf Burgers} equation
\index{Burgers equation}
\begin{equation}
u_t + u_{xx} + (u^2)_x = 0\ :\ u=\varphi_x\varphi^{-1}\ 
\end{equation}
\begin{equation}
u={\cal D}\Log\varphi=\partial_x\Log\varphi\ ;
\varphi_t + \varphi_{xx}+ K(t) \varphi=0,\  
K(t)\hbox{arbitrary function}.
\end{equation}   

{\it Example} 2 (linearisation).
{\bf Liouville} equation
\index{Liouville equation}
\begin{eqnarray}
& &
v_{xt} - e^v = 0
\\
e^v
& = &
u,\
u u_{xt} - u_x u_t - u^3 = 0\ :\ u = 2(\varphi_x\varphi_t\varphi^{-2} -
\varphi_{xt}\varphi^{-1})
\\
u 
& = &
{\cal D} \Log \varphi = - 2 \partial^2_{xt}(\Log \varphi),\
\varphi_{xt}=0.
\end{eqnarray}

{\it Example} 3 (linearisation).
{\bf Eckhaus} equation
\cite{CalogerodeLillo,ClarksonCosgrove1987,CMGallipoli1993}
\index{Eckhaus equation} % Kundu-Eckhaus?
\begin{equation}
\label{eqEckhaus1}
 i u_t + u_{xx} 
+ q_r ({\mod{u}}^4 + 2 a ({\mod{u}}^2)_x) u=0,\
a^2 = 1/q_r,\ q_r \in {\cal R}.
\end{equation}
In the variables $(w, \theta)$ defined by 
$\theta = \arg u, w_x = \mod{u}^2$,
the equation (\ref{eqEckhaus1}) is equivalent to the system
\begin{eqnarray}
\label{eqEckhaus2}
{\hskip -6 truemm}
& &
\theta_x=- {1 \over 2} {w_t \over w_x},\
\theta_t=
 {1 \over 4} (2 {w_{xxx} \over w_x} - {w_{xx}^2 \over w_x^2})
-{1 \over 4} {w_t^2 \over w_x^2}
+ q_r (w_x^2 + 2 a w_{xx})
\label{eqEckhaus3}
\end{eqnarray}
whose compatibility condition is
\begin{eqnarray}
{\hskip -6 truemm}
&\theta_{xt}-\theta_{tx}\equiv&
{\hskip -2 truemm}
(w_{tt} w_x^2 + w_{xx} w_t^2) / 2
+ (w_{xxxx} w_x^2 + w_{xx}^3)/2
- w_x w_{xx} w_{xxx} 
\nonumber
\\
{\hskip -6 truemm}
&\phantom{\theta_{xt}-\theta_{tx}\equiv} &
{\hskip -2 truemm}
- w_t w_x w_{xt}
%\nonumber
%\\
%&\phantom{\theta_{xt}-\theta_{tx}\equiv} &
+ 2 q_r (w_x^4 w_{xx} + a w_x^3 w_{xxx})
=0.
\label{eqEckhaus4}
\end{eqnarray}
Under the transformation $w=(a / 2) \Log \varphi$ defined by the singular part
operator ${\cal D}$ of the equation for $w$,
these three equations become
\begin{eqnarray}
\label{eqEckhaus5}
{\hskip -6 truemm}
& &
\theta_x=- \varphi_t / (2 \varphi_x),\
\theta_t=
 \varphi_{xxx} / (2 \varphi_x)  - \varphi_{xx}^2 / (4 \varphi_x^2)
-\varphi_t^2 / (4 \varphi_x^2)
\label{eqEckhaus6}
\\
{\hskip -6 truemm}
& & 
\theta_{xt}-\theta_{tx}\equiv
(\varphi_{tt} \varphi_x^2 + \varphi_{xx} \varphi_t^2)/2
+ (\varphi_{xxxx} \varphi_x^2 + \varphi_{xx}^3)/2
\nonumber
\\
{\hskip -6 truemm}
& &
\phantom{
\theta_{xt}-\theta_{tx}\equiv
}
- \varphi_x \varphi_{xx} \varphi_{xxx} 
- \varphi_t \varphi_x \varphi_{xt} 
=0.
\label{eqEckhaus7}
\end{eqnarray}
The three equations
(\ref{eqEckhaus5}), (\ref{eqEckhaus6}), (\ref{eqEckhaus7})
are deduced from the three previous ones
(\ref{eqEckhaus2}), (\ref{eqEckhaus3}), (\ref{eqEckhaus4})
by the following simple operation~:
change $w$ to $\varphi$ and assign $q_r$ to zero.
Thus the transformation has linearised the Eckhaus equation
(\ref{eqEckhaus1})
into the Schr\"odinger equation
\begin{equation}
\label{eqEckhaus8}
 i \nu_t + \nu_{xx} = 0,\
\varphi_x = \mod{\nu}^2,\
\arg \nu = \arg u.
\end{equation}
Because of the conservation of the phase, one finally has
\begin{equation}
 u=(\sqrt{a/2}) \nu / \sqrt{\int^x \mod{\nu}^2 \D x},\ 
\mod{u}^2=(a/2) \partial_x\Log\varphi.
\end{equation}

{\it Example} 4 (bilinearisation). 
{\bf \KdV} equation

\index{KdV equation}
\index{bilinear method of Hirota}
\begin{equation}
u_t + u_{xxx} + 3 (u^2)_x = 0\ :\ 
u= -2\varphi_x^2\varphi^{-2} + 2 \varphi_{xx}\varphi^{-1} % ? \'ecrite 2 fois
\end{equation}
\begin{equation}
u = {\cal D}\Log\varphi = 2 \partial_x^2\Log\varphi ;\ 
(D_x D_t + D_x^4) (\varphi\cdot\varphi) = 0
\end{equation}
The transformed equation, quadratic in $\varphi$ 
(see the numerous papers of Hirota 
\cite{Hirota1971,Hirota1974}
for the definition of the bilinear operators $D_x,D_t$) 
is convenient to look for $N-$soliton solutions, 
auto-BTs,
Miura transformations.

% ---------------------------------------------------------------------------
\subsection{Weiss method and its limitations}
\label{sectionWeissLimitations}
\indent

\index{Weiss truncation}

If a nonlinear PDE passing the \Plv\ test is not linearisable,
the idea of Weiss \cite{WTC,Weiss1983} is that the principal part 
of this {\it local} Laurent series
contains all the information for proving that the PDE
possesses the \Plv\ property through the knowledge of its BT
(i.e.~its DT and Lax pair).
This method consists of truncating the Laurent series
for $u$ and $E(u)$
to their nonpositive powers in $\chi$ 
\begin{equation}
\label{eqWeissTruncation} 
u_T=\sum_{j=0}^{-p} u_j \chi^{j+p},\ E_T=\sum_{j=0}^{-q} E_j \chi^{j+q}
\end{equation}
and identifying to zero the coefficients $E_j$ of the 
$\chi-$polynomial $\chi^{-q} E_T(\chi)$.

\noindent Equations $ E_j = 0 $ for $ j = 0,\dots,-p $ determine the $ p+1$ 
coefficients $u_j$ as equal to those of the infinite expansion. 
After replacement of $u_j$ by these values, the remaining equations are
\begin{eqnarray}
\label{eqPD}
&E_j(D\varphi,u_i) = 0&          j\in \{ -p+1,\dots,-q\} \\
&\phantom{E_j(D\varphi,u_i) = 0}&j\ne \hbox{compatible indices },\
i=\hbox{indices}\in\{0,\dots,-p\} \nonumber
\end{eqnarray}
In the Conte gauge, the coefficients $u_j,E_j$ depend on the derivatives of 
$\varphi$ through the homographic invariants $(S,C)$ and their derivatives. 
\index{Riccati equation}
As the variable $\chi^{-1}= \psi_x/\psi$ satisfies a Riccati 
equation one can connect the monomial $(\psi_x/\psi)^n$ with the derivatives
$(\Log\psi)_{jx},(j\le n \in \cal N^{+}) $ and show that
\begin{equation}
\sum_{j=0}^{-p-1} u_j(S,C) (\psi / \psi_x)^{j+p} \equiv
\sum_{j=1}^{-p} \tilde u_j(S,C) (\Log \psi)_{jx} + f(S,C).
\end{equation}
Then the relation
\begin{equation}
\label{eqCDT} 
u_T - \tilde u = \sum_{j=1}^{-p} \tilde u_j(S,C) (\Log \psi)_{jx} 
= {\cal D} \Log \psi,
\end{equation} 
where ${\cal D}$ is the singular part operator and 
$\tilde u=u_{-p}(S,C) + f(S,C)$,  defines a Darboux transformation if 
$E(\tilde u)=0$.
\index{Darboux transformation (DT)}
For this reason, we call equations (\ref{eqPD}) 
  {\it Painlev\'e-Darboux} equations.  
\index{Painlev\'e-Darboux equations}  
The elimination  of the arbitrary functions $u_i$ among this set must produce
only one ``independent" equation  
\begin{equation}
\label{eqSME}
 F(S,C)=0 
\end{equation}
called the {\it singular manifold equation}, 
\index{singular manifold equation}
modulo the ever present link between $S$ and $C$ given by
equation (\ref{eqCrossXT}).

The next step consists of finding a parametric representation for equation
(\ref{eqSME}) under the form $(S,C)$  depending on a function $U$ and an
arbitrary constant $\lambda$ such that the cross-derivative condition 
(\ref{eqCrossXT}) is
identical to the original equation $E(U)=0$ for $U$. 
If this is indeed the case and if $U$ can be identified with $\tilde u$, 
the truncation will provide the DT 
(as a consequence of equation (\ref{eqCDT})), 
the Lax pair 
(as consequences of the linear system
(\ref{eqLinearOrder2X})--(\ref{eqLinearOrder2T}) 
and the parametrisation of $S$ and $C$) 
and thus the BT.
\index{B\"acklund transformation}

This method only succeeds for a few equations, 
like KdV \cite{WTC}, KdV5, AKNS
\index{AKNS equation}
\cite{MusetteSainteAdele}, all belonging to the same hierarchy. 
Let us describe it for the KdV equation (\ref{eqKdVcons}). 
\index{KdV equation}
This equation, which passes the \Plv\ test, admits 
the single family $ u \sim -2 \chi^{-2} $ with Fuchs indices $ -1,4,6 $
\begin{equation}
 u = -2\chi^{-2} + (C-4 S)/6 -(1/6)(C-S)_x \chi + O(\chi^2).
\end{equation}
% This is quite analogous to the proof of `int\'e gration parfaite' (achieved)
% ... through a logarithm derivative ...

The algorithmic results of the \Plv\ analysis for KdV are given in Table 
\ref{Table3.1}.
\index{KdV equation}
They yield
the SM equation
\index{singular manifold equation}
\begin{equation}
C-S + 6\lambda = 0,\ \lambda = \hbox{ arbitrary constant}.
\end{equation}
Its parametric representation 
\begin{equation}
\label{eqSMparametric} 
S=2 (U +   \lambda),\ C=2 (U - 2 \lambda)
\end{equation} 
provides the second order linear system
(\ref{eqLinearOrder2X})--(\ref{eqLinearOrder2T})
\index{Lax pair of second order}
\begin{eqnarray} 
& &
\psi_{xx} + ( U + \lambda) \psi=0
\label{Lax1KdV}
\\ & &
\psi_t + 2 (U - 2 \lambda) \psi_x - U_x \psi=0
\label{Lax2KdV}
\end{eqnarray} 
satisfying the cross-derivative condition
$\psi_{xxt} - \psi_{txx} \equiv 2 \hbox{ KdV}(U) \psi=0$. 
The map between two solutions of KdV coming out of the truncation is
\begin{equation}
u_T =  2 (\Log \psi)_{xx} + (C + 2 S)/6 = 2 (\Log \psi)_{xx} + U
\label{DTKdVcons} 
\end{equation} 
Thus, the Weiss truncation yields both the Lax pair 
(\ref{eqKdVLaxX})--(\ref{eqKdVLaxT}) and the DT (\ref{eqDTKdV}) of the 
KdV equation. 
The auto-BT
(\ref{eqBTKdV1})--(\ref{eqBTKdV2}) is obtained by substitution of the DT
(\ref{DTKdVcons}), i.e.
$\psi_x/\psi = (w-W)/2 $, into the couple (\ref{Lax1KdV}) and (\ref{Lax2KdV})
(notation $u_T=u=w_x, U=W_x$).   
\index{B\"acklund transformation}

It happens that,
for other equations possessing either one family of 
movable singularities or several families with nonopposite residues like 
Boussinesq, Sawada-Kotera, Hirota-Satsuma
\cite{HirotaSatsuma1976} 
equations,
\index{Sawada-Kotera equation}
\index{Hirota-Satsuma equation}  
the parametrisation of $(S,C)$ yields a condition (\ref{eqCrossXT}) for
$U$ different from the original equation, this defines a transformation 
between $u_T$ and $U$, 
\index{Miura transformation}
called {\it Miura transformation}, obtained by the elimination of $(\chi,S,C)$
between the four equations~:
$u_T=$ the truncation, 
the two equations of the parametric representation $(S,C)=f(U)$
and anyone of the two (nonindependent) equations 
(\ref{eqChiX}), (\ref{eqChiT}). 
Then in order to obtain the auto-BT, one requires that the function $\psi$
in equation (\ref{eqCDT}) satisfies a linear third order system whose
coefficients are to be determined as functions of $\lambda$ and another
solution $U$ of the analy-sed PDE linked to $u_T$
through the Darboux transformation.

%-------------------------------NLPDEs with Lax order 3----------------
\subsection{Method for third order Lax pair}
\indent

\index{Lax pair of third order}

%One thus assumes the existence of a Darboux transformation (\ref{eqCDT}) 
%between two different solutions $u_T$ and $\tilde u$ of the same equation
%$E(u)=0$, and one expands $E(u_T)$ as a polynomial in the two independent
% components
%$Z_1 = \psi_x / \psi, Z_2 = \psi_{xx} / \psi$ \cite{MusetteConte1991}. 
Let us denote
$(a,b,c,d,e)$ the five unknown coefficients defining a third order linear 
system for $\psi$
\begin{eqnarray}
\psi_{xxx}  & = & a \psi_x + b \psi
\label{eqLinear3X}
\\
\psi_t  & = & c\psi_{xx} + d \psi_x + e \psi
\label{eqLinear3T}
\end{eqnarray} 
whose compatibility condition is
\begin{equation} 
(\psi_t)_{xxx} - (\psi_{xxx})_t
\equiv X_0 \psi + X_1 \psi_x + X_2 \psi_{xx} = 0
\end{equation}
\begin{eqnarray} 
X_0  & \equiv &
 - b_t - a e_x + e_{xxx} + b_{xx} c 
\nonumber
\\ & & + 3 b c_{xx} + 3 b_x c_x + 3 b d_x + b_x d = 0
\label{eqX0}
\\ X_1  & \equiv &
 - a_t + 3 e_{xx} + 2 b_xc + a_{xx} c + d_{xxx} + 3 a c_{xx}+ 2 a d_x
\\ & & + 3 a_x c_x + 3 b c_x + a_xd = 0
\label{eqX1}
\\ X_2  & \equiv &
 (2 a c + c_{xx} + 3 d_x + 3 e)_x=0.
\label{eqX2}
\end{eqnarray}

In the two independent components
$Z_1 = \psi_x / \psi, Z_2 = \psi_{xx} / \psi$,  
the linear system (\ref{eqLinear3X})--(\ref{eqLinear3T}) is equivalent to the
projective Riccati system \cite{AHW1982}
\index{projective Riccati system}
\begin{eqnarray} 
{\hskip -5 truemm}
Z_{1,x} & = & (- Z_1)Z_1 + Z_2
\label{eqProjRiccatiZ1x}
\\ 
{\hskip -5 truemm}
Z_{2,x} & = & (-Z_1)Z_2 + aZ_1 + b
\label{eqProjRiccatiZ2x}
\\
{\hskip -5 truemm}
 Z_{1,t}  & = & (-dZ_1 -cZ_2)Z_1 + (ac+d_x)Z_1 + (c_x+d)Z_2 +e_x+bc
\label{eqProjRiccatiZ1t}
\\ 
{\hskip -5 truemm}
Z_{2,t} & = & (-dZ_1-cZ_2)Z_2 +(2ac_x+a_xc+bc+d_{xx}+ad+2e_x)Z_1
\nonumber
\\ & &
+(c_{xx}+2d_x+ac)Z_2 +2bc_x+b_xc+bd+e_{xx}.
\label{eqProjRiccatiZ2t}
\end{eqnarray}

The determining equations for the coefficients $(a,b,c,d,e)$ of the Lax pair 
are generated by the expansion of $E_T=E(u_T)$ on the basis $(Z_1,Z_2)$
\begin{eqnarray} E_T & = &
\sum_{l,m} C_{l,m} Z_1^l Z_2^m,
\\ C_{l,m}  & \equiv & C_{l,m}(a,b,c,d,e,U)=0.
\end{eqnarray}

In case the solution of the determining equations does not lead to the
expected solution, for a reason like the absence of a spectral parameter,
the assumption to be changed is the order of the underlying scattering problem.

Let us give more details on the procedure
\cite{MusetteConte1991,MusetteConteGallipoli1992} for finding the BT
of the Boussinesq and Sawada-Kotera equations.
\index{B\"acklund transformation}

%-----------------------------Boussinesq----------------------------------

\subsubsection{First example~: Boussinesq equation} 
\indent

Let us consider the Boussinesq (Bq) equation
\cite{Ursell1953,ZS1974}
\begin{equation}
\label{eqBoussinesqAllConventions}
 E(u) \equiv u_{tt} + \varepsilon^2 \left((u + \alpha)^2 + (\beta^2/3)
u_{xx}\right)_{xx} = 0,
\end{equation}
with $(\alpha,\beta,\varepsilon)$ constant.
% \par We are going to prove that the first 3
%Painlev\'e coefficients 
%$ V_0, V_1 $ and $ V_2 $ are sufficient to obtain the Lax pair.
%As explained for instance in
%section 6.1 of Ref $\lbrack 4 \rbrack$,  
%all the following invariant results can be computed algorithmically~:
% leading term $ u_0 \chi^ p $, i.e.~leading power $ p $ and branch $ u_0 $, 
%indices, sequence of coefficients, set of Painlev\'e-Darboux equations.  
%These are what is 
%usually meant as the output of a PDE Painlev\'e analysis, 
%and it can be  provided by any computer
%algebra program accepting as input the PDE  definition only.  
The algorithmic results of the Painlev\'e analysis are~:
\par
$ p = - 2 $, $ q = - 6 $, indices -1,4,5,6 compatible
\begin{equation}
\label{eqBq4}
u_T =-2\beta^2 \chi^{-2} - 2 \beta^2 S/3 - \varepsilon^{-2} C^2/2,\ 
\chi \hbox{ defined by (\ref{eqChiX})--(\ref{eqChiT})}
\end{equation}
The set of Painlev\'e-Darboux equations reduces to the single equation~:
\index{singular manifold equation}
\begin{equation}
 E_3 \equiv  (1/3) \beta^2 \varepsilon^2 S_x - C_t + C C_x= 0 ,\ 
\end{equation}
which is the SM equation for the Bq equation \cite{Weiss1985Bq} in
the invariant forma-lism. 
This is a conservation law, which can be parametrised as
\begin{equation} 
C =(\beta \varepsilon)^2 z_x,\ 
S = 3 z_t - (3/2)(\beta \varepsilon)^2 z_x^2.
\end{equation}
The compatibility condition of the system 
(\ref{eqLinearOrder2X})--(\ref{eqLinearOrder2T})
reads
\begin{equation}
3 z_{tt} + (\beta \varepsilon)^2 z_{xxxx} + 6 (\beta \varepsilon)^2 z_t z_{xx}
 - 6(\beta \varepsilon)^4 z_x^2 z_{xx} = 0,
\end{equation} 
which is {\it not} the Bq equation but another PDE called modified
Bq equation \cite{HirotaSatsuma1977,FordyGibbons1981}. 
%The result of the truncation
%(\ref{eqBq4}) and the assumption for a
%Darboux transformation (\ref{eqCDT}) writes 
%\begin{equation}
%\tilde u  =  u_T - 2\beta^2 \partial_x^2 \Log \psi + {\beta^2 \over 3}S - 
%{1 \over 2}\varepsilon^{-2}C^2,\
%\hbox{Bq}(\tilde u)=0.
%\end{equation} 
The elimination of $S$ between (\ref{eqBq4}) and (\ref{eqChiX}) yields the
Miura transformation between the Bq and the modified equation
\index{Miura transformation}
\begin{equation}
\label{eqBq5} 
u_T = - (2/3) \beta^2 \chi^{-2} - (1/2) \varepsilon^{-2}C^2
 + (4/3) \beta^2 (\chi^{-1})_x.
\end{equation} 
while the assumption for a DT like (\ref{eqCDT}) leads to
\index{Darboux transformation (DT)}
\begin{equation}
\tilde u =
 - (2/3) \beta^2 \chi^{-2}-(1/2) \varepsilon^{-2}C^2-(2/3) \beta^2 (\chi^{-1})_x,
\end{equation} 
which does not coincide with (\ref{eqBq5}). 
We then conclude that a second order linear system is not convenient 
to represent the Lax pair of the Bq equation.

So, let us assume an underlying scattering problem of the third order for
$\psi$ and the existence of a DT given by the singular part operator
\begin{equation} 
v_T = 2 \beta^2 \Log \psi + V,\ 
\hbox{Bq}(v_{T,xx})=0,\
\hbox{Bq}(V_{xx})=0.
\end{equation} 
Defining the ``second potential Bq'' equation
\begin{equation}
\label{eqppBq} 
F(v) \equiv v_{tt}
+\varepsilon^2\left((v_{xx}+\alpha)^2+(\beta^2/3) v_{xxxx}\right)=0,
% f_1, f_2 \hbox{ arbitrary functions of }t ,
\end{equation} 
% and choosing $f_1=f_2=0$,
$F(v_T)$ is a second degree polynomial in $(Z_1,Z_2)$~:
\begin{equation} 
F(v_T) \equiv
 C_{02} Z_2^2 +C_{11} Z_1   Z_2 +C_{20} Z_1^2 +C_{01} Z_2 +C_{10} Z_1 +C_{00}=0
\label{eqBqZ1Z2}
\end{equation} 
which we identify to zero. This provides
\begin{eqnarray} 
C_{02} & \equiv & 2((\beta \varepsilon)^2-c^2)=0,\ 
\Rightarrow c^2=(\beta \varepsilon)^2
\\ 
C_{11} & \equiv & -4 c d=0,\ 
\Rightarrow d=0
\\ 
C_{20} & \equiv & 
V_{xx}+ \alpha + 2 \beta^2 a/3=0,\ 
\Rightarrow
 a=-3 (V_{xx}+ \alpha)/(2 \beta^2)
\\ 
C_{01} & \equiv &
 2 (\beta \varepsilon^{-1} a c + 2 (V_{xx}+ \alpha) + \beta^2 a/3)=0,\ 
\Rightarrow c=\beta \varepsilon
\\ 
C_{10} & \equiv & 8 (\beta \varepsilon)^2 a_x/3 + 4 e_x c=0,\ 
\Rightarrow e_x=\beta^{-1} \varepsilon V_{xxx},
\\ 
C_{00} & \equiv & 2(\beta^{-2} V_{xxt} +(4/3)\beta \varepsilon b_x + e_{xx}) =0
\\ & &
\Rightarrow 
b=g(t)-(3/4) (\beta^{-2} V_{xxx}+ \beta^{-3} \varepsilon^{-1} V_{xt}).
\end{eqnarray}

Finally, the compatibility condition $X_0=0$ implies that $g(t)$ is an 
arbitrary constant denoted $\lambda$.
The coefficients $a,b,c,d,e$ are  
\index{Lax pair of third order}
\begin{eqnarray} 
& &
a=-(3/2) \beta^{-2}(U+\alpha),\  c=\beta\varepsilon,\ d=0,
\nonumber 
\\ 
& &
b=\lambda 
-(3/4)\beta^{-2} U_x 
-(3/4)\beta^{-3} \varepsilon^{-1} V_{xt},\
e= \beta^{-2} c (U + \alpha),
\end{eqnarray} 
i.~e.~the associated third order Lax pair \cite{Zakharov1973,ZS1974,Morris1976} 
of the derivative of (\ref{eqppBq}) (notation $ U= V_{xx}$).

Since $ d = c_x = 0 $, the BT
obtained by eliminating $ Z_2 $ between 
(\ref{eqProjRiccatiZ1x})--(\ref{eqProjRiccatiZ1t}) 
writes
\index{B\"acklund transformation} 
\index{projective Riccati system}
\begin{eqnarray}
& &
Z_{1,xx} + 3 Z_1 Z_{1,x} + Z_1^3 -a Z_1 -b =0, 
\\
& &
Z_{1,t} + c ( Z_1 Z_{1,x} + Z_1^3 - a Z_1 -\beta^{-2} U_x - b) =0,
\\
& &
(Z_{1,xx})_t - (Z_{1,t})_{xx} = -{3\over 4} \beta^{-3}\varepsilon^{-1} 
(F(V))_x
\end{eqnarray}
or equivalently, with $U=W_x=V_{xx}$ and $Z_1=(w-W)/(2 \beta^2)$,
\begin{eqnarray}
\hspace{-1.5cm}
& &(w-W)_{xx} + 3 \beta^{-2} (w-W)((w+W)_x + 2 \alpha)
 + \beta^{-4} (w-W)^3 
\nonumber 
\\
\hspace{-1.5cm}& &
+ 3 \beta^{-1} \varepsilon^{-1} (w+W)_t -8 \beta^2 \lambda =0,
\\ 
\hspace{-1.5cm}& & 
(w+W)_{xx} + \beta^{-2} (w-W)(w-W)_x - \beta^{-1} \varepsilon^{-1} (w-W)_t =0,
\end{eqnarray}
an extension to $\lambda\not=0$ of the bilinear BT of Hirota and Satsuma
\cite{HirotaSatsuma1977,HirotaSatsuma1978}.
\index{bilinear method of Hirota}

\subsubsection{Second example~: Sawada-Kotera equation}
\indent
  
\index{Sawada-Kotera equation}

In the same way, we can easily find the coefficients of the third order 
Lax pair by processing the fifth order potential equation
\begin{equation}
\label{eqSKpotential}
\hbox{pSK}(v) \equiv
 v_t+v_{5x}+30 v_x v_{3x}+60 v_x^3+F(t)=0,\ F(t) \hbox{ arbitrary} 
\end{equation}

The algorithmic results of the \Plv\ analysis are the following~: 
the equation (\ref{eqSKpotential}) possesses two families,
each with five compatible indices. 
For the ``principal'' family 
\begin{equation}
 p=-1,\ q=-6,\ u_0=1,\ \hbox{indices }-1,1,2,3,10\hbox{ compatible}
\end{equation}
the truncation is
\begin{eqnarray}
& & v_T = \chi^{-1} + v_1. % v_1 \hbox{ arbitrary}
\end{eqnarray}
The assumption $\chi=\psi/ \psi_x$ with $\psi$ solution of the second order 
Lax pair (\ref{eqLinearOrder2X})--(\ref{eqLinearOrder2T})
generates the Painlev\'e-B\"acklund equations \cite{ConteMontpellier89}
\index{Weiss truncation}
\begin{eqnarray}
{\hskip -5 truemm}
E_4 & & \equiv 
 C - 4 S^2 + 9 S_{xx} + 60 S v_{1,x} - 180 v_{1,x}^2 - 30 v_{1,xxx} = 0
\\
{\hskip -5 truemm}
E_5 & & \equiv 
- C_x - 2 S S_x + S_{xxx} + 30 S_x v_{1,x} = 0
\\
{\hskip -5 truemm}
E_6 & & \equiv 
\hbox{pSK}(v_1) + (S E_4 - E_{5,x}) / 2
+ (5/2) S_x (6 v_{1,xx} - S_x) = 0.
\end{eqnarray}
Demanding that $v_1$ be another solution of pSK implies $v_{1,xx}=S_x/6$
and, after computation, provides a nongeneric solution.
Note that, however, a {\it particular} solution of the truncation is
\cite{Weiss1984x}
\begin{eqnarray}
& & S_{xx} + 4 S^2 - C=0,\
v_{1,x}=S/3,\
\hbox{KK}(v_1)=0,
\end{eqnarray}
which defines a Miura transformation between SK and KK equations.
\index{Miura transformation}

As in the preceding example, the hypothesis of the DT
\index{Darboux transformation (DT)}
\begin{equation}
\label{eqDTSKpotential}
v=(\Log\psi)_x+ V,
\end{equation}
with $V$ another solution of pSK and $\psi$ a solution of the 
third order linear system (\ref{eqLinear3X})--(\ref{eqLinear3T}),
makes pSK$(v)-$pSK$(V)$ a second degree polynomial in $(Z_1,Z_2)$ like
(\ref{eqBqZ1Z2}).
The six determining equations $C_{lm}=0$, 
added to the three compatibility conditions (\ref{eqX0})--(\ref{eqX2}),
have the unique solution depending on an arbitrary constant $\lambda$
\index{Lax pair of third order}
\begin{equation}
a=-6V_x,\ b=\lambda,\ c=9\lambda-18V_{xx},\ 
d=-36V_x^2+6V_{3x},\
e_x=36\lambda V_{xx}, 
\end{equation}
a result which coincides with the Lax pair (\ref{eqLax3SK}).

The $x-$part of the BT
(\ref{eqBTSK})
is obtained by eliminating $Z_2$ between (\ref{eqProjRiccatiZ1x}) and
(\ref{eqProjRiccatiZ2x}),
then substituting $Z_1= v - V$ as results from 
(\ref{eqDTSKpotential}).
\index{B\"acklund transformation} 
\index{projective Riccati system}

\subsubsection{Some results for Kaup-Kupershmidt equation}
\indent

\index{Kaup-Kupershmidt equation}
In the case of potential KK equation, the hypothesis of the differential 
operator
${\cal D}=\partial_x$ for the DT, associated to the linear
system (\ref{eqLinear3X})--(\ref{eqLinear3T}),
yields neither the Lax pair (\ref{eqLax3KK})
nor the BT (\ref{eqBTKK}). 
\index{B\"acklund transformation} 
This problem has been recently solved \cite{MusetteConte1998}
by remarking that in his classification of second order first degree nonlinear
ODEs possessing the \Plv\ property,
Gambier \cite{GambierThese} mentions that the following equations~:
\begin{eqnarray}
\hspace{-15mm}& &(G.5)\ :\ 
Y_{1,xx} + 3 Y_1 Y_{1,x} + Y_1^3 + 6 U Y_1 -\lambda = 0 
\label{eqG5}
\\
\hspace{-15mm}& &(G.25)\ :\ 
Y_2 Y_{2,xx} - {3\over 4} Y_{2,x}^2 + {3\over 2} Y_2^2 Y_{2,x} 
+ {1\over 4}Y_2^4 + 6 U Y_2^2 - 2 \lambda Y_2 = 0,
\label{eqG25}
\end{eqnarray}
are linearisable into third order equations % by the transformation
\begin{eqnarray}
\hspace{-3.cm}& &(G.5)\ :\ 
Y_1 = \psi_x / \psi,\ \psi_{xxx} + 6 U \psi_x - \lambda \psi = 0 
\label{eqLinearG5}
\\
\hspace{-3.cm}& &(G.25)\ :\ 
Y_2^{-1} = \lambda^{-1}
[(\psi_x / \psi)_x + (1/2)(\psi_x / \psi)^2 + 3 U], 
\nonumber \\
\hspace{-3.cm}& &\phantom{(G.25)\ :}
\ \psi_{xxx} + 6 U \psi_x + ( 3 U_x  - \lambda) \psi = 0  
\label{eqLinearG25}
\end{eqnarray}
corresponding to the scattering problem of, respectively,
the SK equation for $U$ and
the KK equation for $U$.
It can then be shown that the DTs 
\begin{eqnarray}
Y_1 =    w-W,\  \hbox{with SK}(w_x) =\hbox{SK}(W_x)=0  \\
Y_2 = 2 (w-W),\ \hbox{with KK}(w_x) =\hbox{KK}(W_x)=0 
\end{eqnarray}
leading to the BTs
(\ref{eqBTSK}) and (\ref{eqBTKK}), can be found by singularity analysis. 
\index{B\"acklund transformation}

%------------------------------Two-singular manifold method---------------
\subsection{Two-singular manifold method}
\indent 

For equations with two families of movable poles with opposite residues, 
the truncation procedure
\index{Weiss truncation}
which considers only one family of singularities does not yield the
auto-BT.
\index{B\"acklund transformation}
An extension of the Weiss method consists of considering two distinct 
functions $\psi_1,\psi_2$, assuming now a DT of the form
\begin{equation} 
u_T - U = {\cal D} \Log \psi_1 - {\cal D} \Log \psi_2,\ 
E(u_T)=E(U)=0
\end{equation} 
and, 
assuming that $Y=\psi_1 / \psi_2$ satisfies the most general Riccati system
\index{Riccati equation}
\begin{eqnarray}
\label{eqGeneralRiccatiX1} Y_x & = & R_0 + R_1 Y + R_2 Y^2
\\
\label{eqGeneralRiccatiX2} Y_t & = & S_0 + S_1 Y + S_2 Y^2
\\
\label{eqGeneralRiccatiX3} Y_{xt} - Y_{tx} & \equiv & X_0 + X_1 Y + X_2 Y^2,
\\ X_0 & \equiv &  R_{0,t} - S_{0,x} + R_1 S_0 - R_0 S_1,
\\ X_1 & \equiv & R_{1,t} - S_{1,x} + 2 (R_2 S_0-R_0 S_2),
\\ X_2 & \equiv & R_{2,t} - S_{2,x} + R_2 S_1 - R_1 S_2,
\end{eqnarray} 
eliminating derivatives of $Y$ and identifying $E_T$ to the null polynomial
in $Y$.

The determining equations so generated for the six unknowns $(R_i,S_i)$  
must have a solution such that each $R_i$ is a linear function of $U$
and an arbitrary constant $\lambda$
and such that at least one of the three cross-derivative conditions
$X_i=0$ is identical to the original equation for $U$. 
In such a case, one has found the DT and the Lax pair, 
i.e.~the BT.
\index{B\"acklund transformation}
In the case of two simple poles with constant opposite residues $\pm u_0$ 
and opposite singular part operators $\pm u_0 \partial_x$, 
the truncation
\begin{equation} 
u_T = u_0 (\Log Y)_x + U
\end{equation} 
becomes, by elimination of $Y_x$ from (\ref{eqGeneralRiccatiX1})
\index{Riccati equation}
\begin{equation} 
u_T = u_0 (R_0 Y^{-1} + R_1 + R_2 Y) + U.
\end{equation} 
This represents an extension of the Weiss truncation to the positive powers of
$Y$.
A similar extension was previously made \cite{Pickering1993} in the variable 
$\chi = \psi / \psi_x$ for obtaining particular solutions of nonlinear PDEs.
The ``two-singular manifold'' method is successful for finding the auto-BT
\index{B\"acklund transformation} 
of MKdV and sine-Gordon equations \cite{MusetteConte1994} 
but only partially for the NLS equation.
\index{MKdV equation}
\index{sine-Gordon equation}
\index{nonlinear Schr\"odinger (NLS) equation}
Before detailing this result, let us first reproduce in Table \ref{Table3.1}
the algorithmic results of the \Plv\ analysis for the four
equations belonging to the AKNS scheme\index{AKNS scheme}; 
these include the SM
equation associated with these well known NLPDEs which pass the \Plv\ test.

\index{singular manifold equation}
\index{KdV equation}
\index{MKdV equation}
\index{nonlinear Schr\"odinger (NLS) equation}
\begin{table}[h]
               % Incoh\'erence entre \label et \ref     MODIF
\caption{Algorithmic results of the \Plv\ analysis.
The integers $(p,q)$ are defined in (\ref{eqWeissTruncation}); 
for sG, the polynomial PDE is
(\ref{eqsGPolynomial}). Next column lists the indices, excepted $-1$. 
Column ``PD equations'' lists the subscripts of the non identically zero 
Painlev\'e-Darboux equations; in the sG and NLS
cases, they depend on the arbitrary coefficients introduced at the index $2$ 
(sG) and $0$ (NLS).
}\label{Table3.1}
\vspace{0.4cm}
\begin{tabular}{|| c | l | l | l | l | l ||}
\hline Name &
$p$  &
$q$ & Indices & PD eq.~ & Singular 
\\ &&&&& manifold 
\\ &&&&& equation
\\
\hline KdV &
$-2$  &
$-5$ &
$4,6$ &
$3,5$ &
$S - C=6 \lambda$
%\\
%\hline Bq &
%$-2$  &
%$-6$ &
%$4,5,6$ &
%$3$ &
%$(S_x / 3) - C_t$
%\\ &&&&&
%$+ C C_x = 0$
\\
\hline MKdV &
$-1$  &
$-4$ &
$3,4$ &
$2$ &
$S-C=0$
\\
\hline sG &
$-2$  &
$-6$ &
$2$ &
$3,4,5,6$ &
$S + C^{-1} C_{xx}$
\\ &&&&&
$- C^{-2} C_x^2/2$
\\ &&&&&
$+ 2 \lambda = 0$
\\
\hline NLS &
$(-1,-1)$  &
$(-3,-3)$ &
$0,3,4$ &
$2,2,3$ &
$C_t + 3 C C_x - S_x$
\\ &&&&&
$+ 8 \lambda C_x=0$
\\
\hline
\end{tabular}
\end{table}

Let us now use the information contained in the SME.

\subsubsection{Modified \KdV}
\label{sectionTruncationMKdV}
\index{MKdV equation}
\indent

The equation (\ref{eqMKdVcons}) has two families $u \sim \pm a \chi^{-1}$,
denoted $u \sim a \chi^{-1}$ since $a$ is defined by its square. 
The truncated expansion of a family is
\index{Weiss truncation}
\begin{equation} 
u_T= a \chi^{-1}.
\label{eqMKdV1}
\end{equation} 
The SME $S-C=0$ is parametrised as
\index{singular manifold equation}
\begin{equation} 
S=2 v,\ C=2 v,\ \hbox{KdV}(v) = 0,
\label{eqMKdV2}
\end{equation} 
and the precise relation between $u$ and $v$ (Miura transformation) is 
obtained by eliminating $\chi$ between (\ref{eqMKdV1}) and (\ref{eqChiX})
\index{Miura transformation}
\begin{equation}
(u_T/a)_x + (u_T/a)^2 = - v.
\label{eqMiura2}
\end{equation} 
In fact there are two such Miura transformations, one for each sign of $a$,
i.e.~one for each family.
\index{KdV equation}

Let us first obtain the Darboux transformation for MKdV from that of KdV and 
show that it involves two SMs. 
The Darboux transformation for KdV has been obtained in
section \ref{sectionWeissLimitations}, eq.~(\ref{DTKdVcons}). 
The two Miura transformations (\ref{eqMiura2}) 
and the parametrisation  (\ref{eqMKdV2}) imply~: 
\begin{eqnarray}
- {S_1 \over 2} & = &
\left({u_T \over a}\right)^2 + \left({u_T \over a}\right)_x
 = 2 (\Log \psi_1)_{xx}+ 
\left({U \over a}\right)^2 + \left({U \over a}\right)_x
\\ 
- {S_2 \over 2} & = &
\left({u_T \over a}\right)^2 - \left({u_T \over a}\right)_x
 = 2 (\Log \psi_2)_{xx}+ 
\left({U \over a}\right)^2 - \left({U \over a}\right)_x
\end{eqnarray} 
and the elimination of the nonlinear terms leads to
\begin{equation} 
u_{T,x} = a (\Log (\psi_1/ \psi_2))_{xx} + U_x,
\end{equation} 
which after one integration yields the Darboux transformation for MKdV
\begin{equation}
\label{eqDMKdV} 
u_T = a (\Log (\psi_1/ \psi_2))_{x} + U.
\end{equation}

With this DT, 
the Lax pair is obtained as explained in the introduction of this section.
\index{Riccati equation}
Setting $Y=\psi_1 / \psi_2$ and taking account of 
(\ref{eqGeneralRiccatiX1})--(\ref{eqGeneralRiccatiX2}), 
every derivative of $Y$ can be  replaced by a polynomial in $Y$. 
Consequently, the Darboux transformation (\ref{eqDMKdV}) becomes identical to
\begin{equation} 
u_T= a(R_0 Y^{-1}+R_1+R_2Y)+U
\end{equation} 
and one must identify to zero the polynomial in $Y$
\begin{equation} E(u_T) \equiv E_T=\sum_{j=0}^8 E_j Y^{j-4}.
\end{equation} 
Among the nine Painlev\'e-Darboux equations $E_j=0$, 
only four ($j=1,2,6,7$) are not identically zero.  
Their resolution, as detailed in \cite{MusetteConte1994},
yields the following parametric representation of the six unknowns $R_i,S_i$ 
in which $R_i$ is linear in $U$ and the spectral parameter $\lambda$
\begin{eqnarray}  
R_0 &=& \lambda,\  
R_2=-\lambda,\  
R_1=-2 U/a
\\
S_0 &=& -4 \lambda^2 +2 (U/a)^2 -2 U_x/a,\ 
S_1=8\lambda^2 U/a -4 (U/a)^3 + 2 U_{xx}/a
\nonumber \\  
S_2&=&-4\lambda^2+2 (U/a)^2 +2 U_x/a.
\end{eqnarray} 
This solution associated with the Riccati equations
(\ref{eqGeneralRiccatiX1})--(\ref{eqGeneralRiccatiX2}) reproduces the equations
(\ref{eqMKdVYx})--(\ref{eqMKdVYt}) for the pseudopotential of MKdV .
\index{Riccati equation}

\subsubsection{Sine-Gordon}
\indent

The sine-Gordon equation (\ref{eqsG}), invariant by parity on $u$, is first 
transformed into a
polynomial equation for $v=e^{i u}$, invariant under $v \to 1 / v$.
\index{sine-Gordon equation}
\begin{equation}
\label{eqsGPolynomial}
\hbox{PsG}(v) \equiv 2 v v_{xt} - 2 v_x v_t - v^3 + v =0,\ v=e^{i u}.
\end{equation} 
This PDE has two families of movable singularities
$v=v_1 \sim - 4 C_1 \chi^{-2}$ and $v=v_2^{-1} \sim - 4 C_2 \chi^{-2}$. 
\index{Weiss truncation}
The truncation equations have the
following general solution \cite{Weiss1984c,Conte1989}. 
For the first family ($(v,S,C,\psi)$ are subscripted with $1$)
\index{singular manifold equation}
\begin{eqnarray}
S_1 & = & - v_{1,xx}/v_1 + v_{1,x}^2/(2 v_1^2) - 2 \lambda
 = - i u_{xx} + u_x^2/2 - 2 \lambda
\\ 
C_1 & = & - v_1/ (4 \lambda) = - e^{i u}/ \lambda
\\ 
v_1 & = & - 4 (\Log \psi_1)_{xt} + V_1,\ \hbox{PsG}(V_1)=0.
\label{eqsGCDT1}
\end{eqnarray} 
For the second family $e^{-i u} = v_2 \sim - 4 C_2 \chi^{-2}$
\begin{eqnarray}
S_2 & = & - v_{2,xx}/v_2 + v_{2,x}^2/(2 v_2^2) - 2 \lambda
 =  i u_{xx} + u_x^2/2 - 2 \lambda
\\ 
C_2 & = & - v_2/ (4 \lambda) = - e^{- i u}/ \lambda
\\ 
v_2 & = & - 4 (\Log \psi_2)_{xt} + V_2,\ \hbox{PsG}(V_2)=0.
\label{eqsGCDT2}
\end{eqnarray} 
If one considers only one of these two equivalent SMs, 
the Schwarzian $S_i, i=1 \hbox{ or } 2$, 
does depend on an arbitrary constant $\lambda$ but it has two drawbacks~: 
it is not invariant under parity on $u$, it is not linear in the physical field
$u$ as requested for the Lax pair
(\ref{eqLinearOrder2X})--(\ref{eqLinearOrder2T}) to be a ``good'' one.

Since $v_1-v_2=2 i \sin u$, 
the difference of (\ref{eqsGCDT1}) and (\ref{eqsGCDT2}) reads
\begin{equation}
\sin u = 2 i (\Log (\psi_1/ \psi_2))_{xt} + \sin U,\ 
\hbox{sG}(U)=0,
\end{equation} 
i.e., from the definition of the equation
\begin{equation} 
u_{xt} = 2 i (\Log (\psi_1/ \psi_2))_{xt} + U_{xt}.
\end{equation} 
Integrating twice, we finally obtain the Darboux transformation of sG
\begin{equation}
\label{eqDTsG2} 
u = 2 i \Log (\psi_1/ \psi_2) + U,
\end{equation} 
defined in terms of {\it both} families.
For the solution of the polynomial PDE $\hbox{PsG}(v)=0$ associated to the 
sG equation by $v=e^{iu}$, the DT is~:
\begin{equation} 
v=V Y^{-2},\  Y=\psi_1 / \psi_2,\ 
\hbox{PsG}(V)=0,
\end{equation}  
and one must identify
\begin{equation} 
E(v)=V^2 \sum_{i=0}^4 E_i Y^{i-6}
\end{equation} to the null polynomial in $Y$. 
Among the five Painlev\'e-Darboux equations,
the equation $E_2$ is functionally dependent on $(E_0,E_1)$, 
a consequence of the compatibility of the index $2$. 
Their resolution yields the Riccati pseudopotential
(\ref{eqsGYx})--(\ref{eqsGYt})
\index{Riccati equation}
\begin{eqnarray} 
Y_x  & = &
%\lambda (1 - Y^2) + (V_x/V) Y =
\lambda (1 - Y^2) + i U_x Y,\
\\ 
Y_t & = & 
% (V - Y^2 V^{-1})/(4 \lambda)=
 ((1-Y^2) \cos U + i (1+Y^2) \sin U)/(4 \lambda),\
\label{eqsG987}
\\
(Y_{xt} - Y_{tx})/Y & = &
% V^{-2} \hbox{PsG}(V) = 
\hbox{sG}(U),
\end{eqnarray}
where the $x$-part is now linear in the spectral parameter $\lambda$ and the 
field $U$ associated with the DT (\ref{eqDTsG2}).

%------------------------------Weiss plus homography--------------------------
\subsection{Weiss method plus homography}
\indent 

\index{Weiss method plus homography}

Pickering \cite{Pickering1996} remarks three drawbacks in the previous method
\begin{description}
\item {(i)} 
any explicit relationship is given between the variable $Y\equiv\psi_1/\psi_2$
and the variable $\chi$ of the invariant \Plv\ analysis while the one between
$\chi$ and $\varphi$
is well defined by the homographic transformation \cite{Conte1989}
\begin{equation} 
\chi= {\varphi \over \varphi_x- \varphi_{xx} \varphi /(2 \varphi_x)},
\end{equation}
\item {(ii)} 
the result of the previous truncation for MKdV does not reveal any relationship
between the MKdV and KdV equations as it would be,
\item {(iii)} 
the knowledge of the DT is required in advance.
\end{description}
He notices that for finding the BT of MKdV and sine-Gordon equations
it is sufficient to consider a  Riccati system  constructed from the
nonlinearisation of the following second order scalar linear system 
\index{B\"acklund transformation}
\index{MKdV equation}
\index{sine-Gordon equation}
\index{Lax pair of second order}
\index{Riccati equation}
\begin{eqnarray}
& & 
\eta_{xx} = 2A \eta_x + B \eta
\label{eqGLinear2x}
\\
& & 
\eta_t = - C \eta_x + \left( \int^x D\ \D x'\right)\eta
\label{eqGLinear2t}
\end{eqnarray}
by the transformation $ Z^{-1} = \eta_x/\eta$. 
The corresponding nonlinear system
\begin{eqnarray}
& & 
Z_x = 1- 2 A -B Z^2
\label{eqNRiccatix} 
\\                                
& & 
Z_t = - C + (C_x + 2 AC)  Z - (D - BC) Z^2
\label{eqNRiccatit}
\end{eqnarray}
depends on four functions $ A,B,C,D$ 
in place on six like the system
(\ref{eqGeneralRiccatiX1})--(\ref{eqGeneralRiccatiX2}) considered in the 2-SM 
method. 
Its compatibility condition is
\begin{eqnarray}
& & Z_{xt}-Z_{tx} \equiv X_1 Z + X_2 Z^2 = 0,\ 
\\
& & X_1 \equiv 2 \left( D - (A_t + ( AC )_x + C_{xx}/2)\right) = 0
\label{Pickeringcc1}  
\\
& & X_2 \equiv D_x - B_t - 2 B C_x - B_x C - 2AD = 0
\label{Pickeringcc2}
\end{eqnarray}
The solution $\eta$ of the linear ODE (\ref{eqGLinear2x}) is related to the 
solution $\psi$ of
(\ref{eqLinearOrder2X}) by the gauge transformation
\begin{equation} 
 \eta=\left( e^{\int^x A\ \petD x'}\right) \psi
\end{equation}   
with
\begin{equation}
\label{eqSAB}
 S=-2( B+ A^2 - A_x )
\end{equation}
Then, computing the $x-$derivative of $\Log \eta$, one gets the transformation
\begin{equation}
\label{Zvar}
Z^{-1} = \chi^{-1} + A 
\end{equation}
which means that the new expansion variable $Z$ is related to $\chi$ by an 
homographic transformation 
(as suggested in \cite{MusetteConte1991}, formula (17)) 
such that in the neighbourhood of $\chi = 0$, one has $Z\sim \chi$.
Then the system (\ref{eqNRiccatix})--(\ref{eqNRiccatit}) combined with the
truncation in $Z$
\begin{equation}
\label{newtrun}
u_T = a \partial_x\Log Z + U ,\  
\end{equation}
($U$  function of ($x,t$) and $a$ constant) such that 
$ u_T\sim a \chi^{-1} $ as $ \chi \to 0 $,
 extends the Weiss truncation to positive powers of $Z$.
\index{Weiss truncation}
Solving the Painlev\'e-Darboux equations associated with $E(u_T) = 0$
Pickering obtains for the MKdV equation (\ref{eqMKdVcons}) 
the following results~:
\begin{eqnarray}
& &
A = U/a,\ 
B = \lambda^2,\
C = 2 (U_x/a - (U/a)^2 + 2 \lambda^2)
\label{eqABC}
\\
& & 
D = 4 \lambda^2 U_x/a,\ 
\lambda = \lambda (t) \hbox{ arbitrary integration function}
\end{eqnarray}
and the compatibility conditions (\ref{Pickeringcc1})--(\ref{Pickeringcc2}) 
yield
\begin{eqnarray}
X_1&\equiv&-(2/a) (U_t + U_{xxx} - 2 a^{-2} (U^3)_x) = 0
\\
X_2&\equiv&- (\lambda^2)_t = 0
\end{eqnarray}
From (\ref{eqSAB}) and (\ref{eqABC}) one gets 
\begin{eqnarray}
& &
S =  2 (U_x/a - (U/a)^2 - \lambda^2)
\\
& &
S - C + 6 \lambda^2 = 0,
\end{eqnarray}
\index{singular manifold equation}
the latter equation being the SM equation of the KdV equation. 
\index{KdV equation}

Let us remark that the expressions of $S$ and $C$ in function of the solution
$U$ of the MKdV equation and the constant parameter $\lambda$ 
coincide with the relation (\ref{eqSCMKdV}) obtained previously. 
The expression (\ref{eqY}) for $Y$ with $\alpha = U/a$ implies the 
identification $ Z=\lambda^{-1}Y$.

For the sine-Gordon equation (\ref{eqsG}), considering the truncation
\index{sine-Gordon equation}
\begin{equation}
   u_T = 2i \Log Z + U
\end{equation}
such that $ u_T \sim 2i\Log\chi$ as $\chi \to 0$, Pickering obtains~: 
\index{singular manifold equation}
\begin{eqnarray} 
{\hskip -16 truemm}
& &
A = -(i/2) U_x,\ 
B = \lambda^2,\ 
D = -(i/2) \sin U,\
\lambda\hbox{ arbitrary constant}, 
\label{eqABD}
\\ 
{\hskip -16 truemm}
& & 
C = - \lambda^{-2} e^{iU}/4,\
S = -iU_{xx} + (1/2) U_x^2- 2 \lambda^2,\ 
\hbox{sG}(U)=0,
\\
{\hskip -16 truemm}
& &
S + C_{xx}/C- (1/2) (C_x/C)^2 + 2 \lambda^2 = 0
\label{SMsG}
\end{eqnarray}
Again the expressions for $S$, $C$  in function of $U$ and $\lambda$ and the 
equation (\ref{SMsG}) coincide with the relations (\ref{eqSCsineG}) and 
(\ref{SMsineG}). 
The identification $ Z=\lambda^{-1}Y$ is also easy to find taking account that,
for sG, $\alpha = -i u_x/2$.

%-----------------------------Weiss+point symmetry (AKNS system) ------------

\subsection{Weiss method plus involutions}
\indent 

\index{Weiss method plus involutions}
The AKNS system~\cite{ZS1971,AKNS}\index{AKNS system}
\begin{equation} 
E^{(1)} \equiv   i u_t + p_r u_{xx} + q_r u^2 v = 0,\ E^{(2)} \equiv - i v_t +
p_r v_{xx} + q_r u v^2 = 0
\label{eqAKNS}
\end{equation} has the BT
\cite{Lamb1974,Chen1974,KonnoWadati1975,LeviRagniscoSym1984} %NeugebauerMeinel
$(a^2=-2 p_r/q_r,\ R^2=(u+U) (v+V)/a^2 - (\lambda - \mu)^2)$
\index{B\"acklund transformation}
\begin{eqnarray} & (u+U)_x & = - (u-U) R - i (\lambda + \mu) (u+U)
\nonumber
\\ & (v+V)_x & = - (v-V) R + i (\lambda + \mu) (v+V) 
\nonumber
\\ & +i p_r^{-1} (u+U)_t & = (u-U)_x R + (u+U) M + i (\lambda + \mu) (u+U)_x
\nonumber
\\ & -i p_r^{-1} (v+V)_t & = (v-V)_x R + (v+V) M - i (\lambda + \mu) (v+V)_x
\nonumber
\\ & M & = (u v + U V)/a^2 % = [(u+U)(v+V)+(u-U)(v-V)]/(2 a^2)
\label{eqAKNSBT}
\end{eqnarray}
with $\lambda,\mu$ arbitrary complex constants. 
Galilean invariance
$(x,t,u,v)$
$\to (x- 2 p_r c t,t,e^{i (c x - p_r c^2 t)}u, e^{-i (c x - p_r c^2 t)}v)$ 
allows to choose
$c=\lambda + \mu=0$~\cite{KonnoWadati1975}.

The above BT
\index{B\"acklund transformation} 
cannot be found neither by the one--SM method~\cite{Weiss1985Bq}, 
nor by the two--SM method~\cite{MusetteConte1994}, 
nor by the one--SM method plus homography \cite{Pickering1996}. 
The challenge of the Painlev\'e approach to find this BT
by singularity analysis {\it only} is solved in \cite{CMGalli95} as follows.
\index{B\"acklund transformation} 

As the one--SM method only provides some partial result
$T(\chi,u,\lambda)$ for the truncation, 
\index{Weiss truncation}
one then considers all transformations on $u$ conserving
the equation
$E(u)=0$ in order to uncover a second solution $U$,
see Table \ref{Table3.2}.

\index{MKdV equation}
              % Incoh\'erence entre \label et \ref     MODIF
\begin{table}[h]
\caption{
Transformations of the dependent variable(s) conserving
the equation(s), 
for the AKNS group PDEs
(complex conjugation, phase shift, parity).
}\label{Table3.2}
\vspace{0.4cm}
\begin{center}
\begin{tabular}{| l | l |}
\hline
PDE
&
Transformation(s)
\\ \hline
AKNS system\index{AKNS system}
&
$
(u,v,i) \to (v,u,-i);\
\forall k:\ (u,v) \to (k u, v/k)
$
\\ \hline
\index{sine-Gordon equation}
Sine-Gordon
&
$u \to -u$ 
\\ \hline
MKdV
&
$u \to -u$
\\ \hline
KdV
&
none
\\ \hline
\end{tabular}
\end{center}
\end{table}

For the AKNS system\index{AKNS system} (\ref{eqAKNS}), 
the one--family truncation
\index{Weiss truncation}
\begin{equation}
u=u_0 \chi^{-1} + u_1,\
v=v_0 \chi^{-1} + v_1
\end{equation}
which has the general solution~\cite{Weiss1985Bq,MusetteConte1994}
($\lambda$ arbitrary complex constant)
\begin{eqnarray}
{\hskip -13 truemm}
& &u
= 
a (\chi^{-1} - f_x/(2f) - i \lambda) f
\label{eqTu}
\\ 
{\hskip -13 truemm}
& &v
= 
a (\chi^{-1} + f_x/(2f) + i \lambda)/f
\label{eqTv}
\\
{\hskip -13 truemm}
%(\Log f)_x
& &f_x/f
=
-2 i \lambda - (u / a) f^{-1} + (v / a) f
\label{eqTx}
\\
{\hskip -13 truemm}
%i p_r^{-1} (\Log f)_t
& &i p_r^{-1} f_t/f
=
2 u v / a^2 + 4 \lambda^2
%+ ((u_x - 2 i \lambda u) / a) f^{-1}
 +  (u_x - 2 i \lambda u) / (a f)
%+ ((v_x + 2 i \lambda v) / a) f
 +  (v_x + 2 i \lambda v) f /a
\label{eqTt}
\\
{\hskip -13 truemm}
%(\Log f)_{xt} - (\Log f)_{tx}
& &(f_{xt} - f_{tx})/f
=
(f^{-1} E^{(1)} + f E^{(2)}) / a
\label{eqTxt}
\end{eqnarray}
fails to introduce a second solution $(U,V)$, 
see details in appendix C of Ref.~\cite{MusetteConte1994}.
This is done by applying the two point transformations of Table 3.2 to 
the above truncation $T_1$ (\ref{eqTu})--(\ref{eqTt})~:
\begin{equation}
\left. \begin{array}{rlrrrlll}
T_1:&\chi_1&       u &       v &  i & f & \lambda  & \hbox{(identity)}    \\
T_2:&\chi_2&       v &       u & -i & g & \mu      & \hbox{(conjugation)} \\
T_3:&\chi_3& k     U & k^{-1}V &  i & f & \lambda' & \hbox{(phase shift)} \\
T_4:&\chi_4& k^{-1}V & k     U & -i & g & \mu'     & \hbox{(both)}
\end{array} \right\}
\end{equation}
These transformations act on $(u,v,f,\lambda)$ like in Chen~\cite{Chen1974}.
This is equivalent to successively process the four families of the AKNS
system\index{AKNS system} by the one--SM method.
In order that $(u,v)$ and $(k U, V/k)$ be distinct,
one must have $\lambda'=\mu,\mu'=\lambda$.

The four sets (\ref{eqTu})--(\ref{eqTv}) define a system of
eight equations in the eight unknowns
$(\chi_1^{-1},\chi_2^{-1},\chi_3^{-1},\chi_4^{-1},u,v,k U,V/k)$.
This system is linear with determinant $f g -1/(f g)$
and it provides the DT straightforwardly
(with the nonrestrictive choice $k=-1$)~:
\begin{eqnarray}
u-U
& = & 
 2 a[\partial_x \Log(g-1/f) - i (\lambda + \mu)]/(g+1/f)
\label{eqDTAKNSdiffu}
\\
v-V
 & = &
 2 a[\partial_x \Log(f-1/g) + i (\lambda + \mu)]/(f+1/g)
\label{eqDTAKNSdiffv}
\\ 
u+U
& = & 
 2 i a (\lambda - \mu)/(g-1/f)
\label{eqDTAKNSsumu}
\\
v+V
& = &
 2 i a (\lambda - \mu)/(f-1/g)
\label{eqDTAKNSsumv}
\end{eqnarray}
(to stick to our definition, the DT is made of two equations,
either (\ref{eqDTAKNSdiffu})--(\ref{eqDTAKNSdiffv}) or 
(\ref{eqDTAKNSsumu})--(\ref{eqDTAKNSsumv})).
The nonconstant factor of the logarithmic derivatives
is similar to that of (P3), (P5), (P6), 
see section 7.1 in the Conte lecture                                 % MODIF
while $\lambda + \mu$ is choosen as a real constant. 

\index{Riccati equation}
The Lax pair in its Riccati form
is made of the four equations resulting from the action of
$T_3$ and $T_4$ on (\ref{eqTx})--(\ref{eqTt}).
The BT
\index{B\"acklund transformation} 
is made of the four equations resulting from the
elimination of  the pseudopotentials $(f,g)$
between the six equations defining the DT and the Lax pair,
and these are precisely (\ref{eqAKNSBT}).
This elimination is quite easy since equations (\ref{eqDTAKNSsumu})--(\ref{eqDTAKNSsumv}) 
are algebraic in $(f,g)$~:
\begin{eqnarray}
{\hskip -10 truemm}
& &
f= i a (\lambda - \mu + R)/(v+V),\
g= i a (\lambda - \mu + R)/(u+U).
\end{eqnarray}

{\it Remark}.
The system \cite{Chen1974} of two equations
for % the two Riccati pseudopotentials 
$(f,g)$,
obtained by eliminating $(u,v,U,V)$ 
between (\ref{eqDTAKNSdiffu})--(\ref{eqDTAKNSsumv}) and the PDE,
is invariant under $(\lambda,\mu) \to (\mu,\lambda)$.
The elimination of $g$ between this system
provides the Broer-Kaup equation for $w=- i \Log f$,
a result also obtainable by the Weiss truncation~\cite{MusetteConte1994}
\index{Weiss truncation}
\index{Broer-Kaup equation}
\begin{eqnarray}
& &
p_r^{-1} w_{tt} + 4 w_x w_{xt} + 2 w_t w_{xx} + p_r (6 w_x^2 w_{xx} + w_{xxxx})
=0.
\end{eqnarray}

% =========================== Reductions of the DT of AKNS system ============

\subsection{Reductions of the DT of AKNS system\index{AKNS system}}
\indent

The $x-$part of the AKNS spectral problem\index{AKNS scheme} admits the 
three reductions
$v=\overline{u}, v = \pm u, v=1$,
and the DT, obtained only from the $x-$part, must admit them.
This is indeed the case~:
equations (\ref{eqDTAKNSdiffu})--(\ref{eqDTAKNSsumv}) admit the two reductions
$(v,V,g,\mu)=
(\overline{u}, \overline{U}, \overline{f}, \overline{\lambda}),
(\varepsilon u,\varepsilon U,\varepsilon f, - \lambda),
\varepsilon^2=1$,
and one must add the case $g=\varepsilon/f$ when the determinant vanishes.
Table \ref{Table3.3}
summarises these reductions and the homographic link between
$f$ and the $\chi$ of the invariant analysis.

\newpage

%\underbar{Table 3}.

\index{nonlinear Schr\"odinger (NLS) equation}
\index{sine-Gordon equation}
\index{MKdV equation}
\index{KdV equation}
\begin{table}[h]
             % Incoh\'erence entre \label et \ref     MODIF
\caption{
Reductions of the Darboux transformation of the AKNS system\index{AKNS system}.
}\label{Table3.3}
\vspace{0.4cm}
\begin{center}
\begin{tabular}{| l | l | l | l | l |}
\hline
PDE
&
$v,g,\mu$
&
$\chi^{-1}$
&
$(u-U)/a$
&
$(u+U)/a$
\\ \hline
% & & & & \\ % to force visibility for \overline
NLS
&
$\overline{u}, \overline{f}, \overline{\lambda}$
&
% N.A. % Is it hopeless?
&
%$2a {\partial_x \Log(\overline{f}-1/f)
% -2 i \hbox{Re } {\lambda} \over \overline{f}+1/f}$
(\ref{eqDTAKNSdiffu}) % to avoid Overfull \hbox 21pt
&
%$- 4 {\hbox{Im }{\lambda} \over \overline{f} - 1/f}$
%$- 4 (\hbox{Im }{\lambda}) / (\overline{f} - 1/f)$
 $  4 (\hbox{Im }{\lambda}) / (1/f - \overline{f})$
\\ \hline
$\displaystyle \hbox{sG} \hfill \atop \displaystyle \vstrut \hbox{MKdV} \hfill$
&
$
\displaystyle
\varepsilon u,\varepsilon f, - \lambda
\hfill
\atop
\displaystyle
e^2=\varepsilon
\hfill
$
&
$
\displaystyle
%\lambda Y^{-1} - e U /(4 a),\
{\lambda \over Y} - {e U \over 4 a},
\hfill
%                                   CHANGE OF NOTATION~: (Y,e) \to (1/Y, -e)
\atop
\displaystyle
Y
{\hskip -0.5truemm}
=
{\hskip -0.5truemm}
{e f - 1 \over e f + 1}
\hfill
$
&
$
\displaystyle
(4 /e) {Y_x \over Y}=
\hfill
\atop
\displaystyle
%2 f_x /(\varepsilon f^2 -1)
{2 f_x \over \varepsilon f^2 -1}
\hfill
$
&
$
\displaystyle
%(i \lambda/e) (1/Y - Y)
%(i \lambda/e) (Y^{-1} - Y)
% i \lambda e^{-1} ({1 \over Y} - Y)
 {i \lambda \over e} ({1 \over Y} - Y)
\hfill
\atop
\displaystyle
= 4 i \lambda {f \over \varepsilon f^2 -1}
\hfill
$
\\ \hline
KdV
&
$1, \varepsilon/f, - \lambda$
&
$
f - i \lambda
$
&
$
\displaystyle
-2 f_x=
\hfill
\atop
\displaystyle
-2 (\chi^{-1})_x
\hfill
$
&
$
\displaystyle
2 (f^2 - 2 i \lambda f) 
\hfill
\atop
\displaystyle
= 2 (\chi^{-2} + \lambda^2)
\hfill
$
\\ \hline
\end{tabular}
\end{center}
\end{table}

% ====================================================================
\section{Cosgrove classification for semilinear PDEs of second order}
\indent

In two papers \cite{CosgroveI1993,CosgroveII1993}, 
Cosgrove classifies two cases of \Plv\ type semilinear PDEs of second order.
The necessary conditions for the \Plv\ property which he establishes
combine the criteria of \Plv\ and Gambier for ODEs and the WTC ones for PDEs.
\index{Painlev\'e test for PDEs}

 For {\bf hyperbolic PDEs} in two independent variables of the type
\begin{equation}
\label{eqHyperbolic}
    u_{xt} = F(x,t,u,u_x,u_t)
\end{equation}
the classes of equivalence are defined by the H-transformation
\begin{eqnarray}
\label{eqHT} 
\tilde{u} &=& {\alpha(x,t) u + \beta(x,t) \over \gamma(x,t) u + \delta(x,t)} 
\\
\hbox{where  } u &=& u(\tilde x,\tilde t),\ 
\tilde x =X(x),\ \tilde t = T(t),\
\alpha\delta - \beta\gamma \ne 0.
\end{eqnarray}
and the necessary conditions are
\begin{enumerate}
\item 
the dependence in $u_x,u_t$ of $F$ must be of the form
\begin{equation}
 u_{xt} = A(x,t,u) u_x u_t + B(x,t,u) u_x + C(x,t,u) u_t + D(x,t,u),
\end{equation}
\item as a function of $u$, the term $A$ is the sum of at most
three simple poles, at locations set to $ u = 0,1,H(x,t)$ 
($H$ arbitrary function of $x,t$) 
while $B,C,D$ cannot grow faster than, respectively, $u,u,u^3$,
\item 
the equation must pass the WTCK \Plv\ test, i.e.~all Fuchs indices are
distinct integer and all positive indices are compatible in order to 
guarantee the existence of local Laurent series in the Kruskal variable 
$\varphi(x,t) = x \pm f_1(t) $ or
$\varphi(x,t) = t
\pm f_2(x) $.
\end{enumerate}
At the end, he obtains 22 canonical equations, reducible to
\index{sine-Gordon equation}
\index{Tzitz\'eica equation}
\begin{equation}
 u_{xt} = \sin u\hbox{  (sine-Gordon) or }\ u_{xt} = a e^{2u} + b e^{-u}\hbox{ 
(Tzitz\'eica)}
\end{equation}     
or linearisable by the {\it singular part} transformation.

For {\bf parabolic PDEs} the class of equivalence is a little bit larger than
in the previous case in the sense that the new independent variables in the
transformation (\ref{eqHT}) may be related to $(x,t)$ as
\begin{equation}
\tilde x = X(x,t),\ 
\tilde t = T(t)
\end{equation}
The successive necessary conditions for having the \Plv\ property yield the
following results
\begin{enumerate}
\item in two independent variables, the sole equation is
\begin{equation}
 u_t + u_{xx} + 2 u u_x = F(x,t),\ 
F(x,t)\hbox{ arbitrary function}
\end{equation}
i.e.~the Forsyth-Burgers equation linearisable into the heat equation,
\index{Burgers equation}
\item in more than two independent variables, only trivial soliton equation are
obtained, i.e.~nonlinear PDEs related to linear ones.
\end{enumerate} 

% ==========================================================================
\section{PDEs with variable coefficients}
\indent

\index{Painlev\'e test for PDEs}
The \Plv\ test can be applied to nonlinear PDEs with variable coefficients in
order to determine the conditions under which the equation might be 
integrable.
The sufficient part of the analysis entails the determination of the DT and 
the Lax pair,
as well as
the transformation which could relate the equation to 
its autonomous integrable counterpart. 
Many authors have considered the generalised KdV and NLS equations with 
variable coefficients due to their interest in many physical systems.
The results of Brugarino \cite{Brugarino1989}
for the generalised variable coefficient KdV equation (VCKdV)
\index{variable coefficient KdV}
\begin{equation}
\label{eqvarKdV}
u_t + a(t) u + \left(b(x,t) u \right)_x +c(t) u u_x +d(t) u_{xxx} +e(x,t) = 0,
\end{equation}
are
\begin{description}
\item (i) 
% J'EN SUIS LA
the equation passes the \Plv\ test under the condition
\begin{eqnarray}
{\hskip -10 truemm}
& & b_t + (a - Lc) b + b b_x + d b_{xxx} = 2a h + h L(d/c^2) + h' + ce
\nonumber
\\ 
{\hskip -10 truemm}
& & 
+ x\left(2 a^2 + a L(d^3/c^4) + a' + L(d/c) L(d/c^2) + (L(d/c))' \right)
\label{varKdVCondition}
\end{eqnarray}
with $L=(\D /\D t)\Log$ and $h(t)$ arbitrary, 
\index{Weiss truncation}
\item (ii) 
the solution of the Weiss truncation yields the DT and the Lax pair,
\item(iii) 
with the transformation from $(u,x,t)$ to $(\Theta,\xi,t)$
\begin{equation}
u=\left((a + L(d/c))\xi + g - b + \Theta \right)/c,\
\xi = x -\int^t g(T) \D T
\end{equation}
and the condition (\ref{varKdVCondition}), 
he gets the equation
\begin{equation}
\Theta_t 
+ \left(2a + L(d/c^2)\right)\Theta + \xi\left(a + L(d/c)\right)\Theta_{\xi} 
+ \Theta\Theta_{\xi}  + d \Theta_{\xi\xi\xi} = 0,
\end{equation}
equivalent to \cite{Hirota1979,BrugarinoPantano1980,Hlavaty1986}
the KdV equation with constant coefficients.
In case  $ a = b = e \equiv 0 $, the
condition (\ref{varKdVCondition}) becomes simply
\index{Painlev\'e test for PDEs}
\begin{equation}
d=c \left( K_1 \int^t c(T) \D T + K_2\right),\
K_1,K_2 \hbox{ arbitrary constants},
\end{equation}
i.e.~the one given by Joshi \cite{Joshi1987} 
when performing the \Plv\ test on this particular VCKdV equation. 
The same relation was obtained by Winternitz and Gazeau 
\cite{WinternitzGazeau1992} using the symmetry group,
\item (iiii) 
several equations of physical interest which satisfy (\ref{varKdVCondition})
are presented.
\end{description}

Gagnon and Winternitz \cite{GagnonWinternitz1993} have analysed from the point
of view of symmetries a variable coefficient nonlinear Schr\"odinger 
(VCNLS) equation 
\index{variable coefficient NLS}
\begin{equation}
\label{eqVCNLS}
 i u_t + f(x,t) u_{xx} + g(x,t) u {\vert u\vert}^2 + h(x,t) u = 0
\end{equation}
involving the three complex functions $f,g$ and $h$ of the variables $x,t$.
The symmetry group is shown to be five-dimensional
iff the equation (\ref{eqVCNLS}) is equivalent to NLS itself or to CGL3,
and at most four-dimensional in all other cases.
\index{complex Ginzburg-Landau (CGL3) equation}
In this framework, 
they give the allowed transformations of (\ref{eqVCNLS}),
in case $ f=1+i f_2$,
to the equation with
constant coefficients
\begin{equation}
\label{eqGCGL3}
 i\tilde u_{\tilde t} 
+ (1+i f_2) \tilde u_{\tilde x \tilde x} 
+ (\tilde g_1 + i \tilde g_2)\tilde u {\vert\tilde u\vert}^2 
+ (\tilde h_1 + i \tilde h_2) \tilde u = 0
\end{equation}
which are, in the special case $f_2 = 0 $
($I,J,K,L$ arbitrary functions of $t$),
\begin{eqnarray}
& & 
{\hskip -5 truemm}
g=(\tilde g_1 + i \tilde g_2) \dot T I^{-2}  
\\
& &
{\hskip -5 truemm}
\Re h = (\dot K + 4 K^2) x^2 + (\dot L + 4 K L)x + \dot J + L^2 
+\tilde h_1 \dot T 
\\
& &
{\hskip -5 truemm}
\Im h = -\dot I I^{-1} -2 K + \tilde h_2 \dot T  
\\
& &
{\hskip -5 truemm}
\tilde t = T,\
\dot T =T_0 e^{-8\int K  \petD t},\   
\tilde x = \sqrt{\dot T} x + \xi,\
\dot\xi= -2\sqrt{\dot T} L 
\\
& &
{\hskip -5 truemm}
u= \tilde u(\tilde x,\tilde t) I e^{i(K x^2 + Lx + J)}.
\end{eqnarray}

One example given by the authors leading to the NLS equation 
($\tilde g_2 = \tilde h_1 = \tilde h_2 = 0 $) 
corresponds to the following choice of the arbitrary functions involved
in the transformation
\begin{equation}
\dot I I^{-1} = -2 K,\ J=L=0,\ \dot K + 4 K^2 = K_0/4,\ K_0\hbox{ constant} 
\end{equation}
and yields the VCNLS equation
\begin{equation}
i u_t + u_{xx} + \tilde g_1 T_0 \sech (\sqrt{K_0} t) u {\vert u \vert}^2 
+ (K_0/4) x^2 u
= 0.
\end{equation}
This equation is related to NLS by the change of variables
\index{nonlinear Schr\"odinger (NLS) equation}
\begin{eqnarray}
& &\tilde x = \sqrt{T_0} x \sech(\sqrt{K_0}t) + x_0,\
   \tilde t =       T_0  \tanh(\sqrt{K_0}t)/\sqrt{K_0}  
\\
& &u(x,t) = \tilde u(\tilde x, \tilde t) \sech^{1/2}(\sqrt{K_0}t)\ 
e^{(1/4) i \sqrt{K_0} x^2 \tanh(\sqrt{K_0}t)}
\end{eqnarray}
Considering the other choice~: $ \dot T I^{-2} = 1$, one obtains
\begin{equation}
\label{eqVCNLS1}
i u_t + u_{xx} + \tilde g_1 u {\vert u \vert}^2 
+ \left((4K^2 + \dot K) x^2 + (\dot L + 4 KL) x + L^2 + \dot J + 2 i K \right)
 u = 0
\end{equation} 
equivalent to NLS by the transformation
\begin{eqnarray}
& &\tilde t= T_0 \int^t \D t' e^{-8 \int^{t'} K \petD s},\ 
\tilde x = x \sqrt{T_0} e^{-4\int^t K \petD t'} + \xi 
\nonumber
\\
& & \xi = \xi_0  - 2\sqrt{T_0}\int^t \D t' L(t') e^{-4\int^{t'} K(s) \petD s} 
\label{eqVCNLST1}
\\
& & u(x,t) = \tilde u(\tilde x,\tilde t)
 e^{-4\int^t K \petD t'}e^{i( K x^2 + L x + J)}.
\nonumber
\end{eqnarray}
For $ K(t) = -\beta (t)/ 2 $, Equ.~(\ref{eqVCNLS1}) and the transformation
(\ref{eqVCNLST1})
coincide with the results of Clarkson \cite{Clarkson1988} 
in the analysis of the PDE
\begin{equation}
  iu_t + u_{xx} - 2 u {\vert u \vert}^2 = a(x,t) u + b(x,t)
\end{equation}
which passes the \Plv\ test iff 
there exist functions $(\beta(t),\alpha_1(t),\alpha_0(t))$ such that
\begin{equation}
\label{VCNLScondition}
a=i \beta + x^2 [\beta'/2 -\beta^2] + x \alpha_1 + \alpha_0,\
b=0.
\end{equation}
As noticed by Clarkson, this result proves that the equation 
\begin{equation}
 iu_t + u_{xx} - 2 u {\vert u \vert}^2 = \beta x^2 u,\ 
\beta\hbox{ constant},\
\beta\not=0,
\end{equation}
which does not satisfy the condition (\ref{VCNLScondition}), 
is not integrable while 
\begin{equation}
iu_t + u_{xx} - 2 u {\vert u \vert}^2 = (i \kappa - \kappa^2 x^2) u,\
\kappa \hbox{ real},
\end{equation}
can be transformed into NLS and hence is integrable. 
 
%\vfill \eject

% ---------------------------------------------------------------------------
% ==========================================================================
\chapter{Partially and non integrable equations}
\label{chap4}
\indent

\index{Painlev\'e test for PDEs}
Let us consider a polynomial PDE (depending on a parameter $\mu$) which does
not pass the \Plv\ test but possesses a singularity structure compatible with
singlevalued solutions. 
The reason for the failure of the \Plv\ test are the following ones~:
\begin{description}
\item (i) 
the Fuchs indices are \underbar{all} integers 
(possibly for particular values  of $\mu$) 
but some of them do not satisfy the compatibility condition,
therefore the existence of a Laurent series is submitted to conditions on the 
SM,
\item (ii) 
whatever be $\mu$,
some indices are irrational and  
thus unable to generate compatibility conditions,
therefore the number of arbitrary functions which can be introduced in the 
Laurent series is lower than the order of the equation; 
this generally characterises equations with chaotic behaviour.
\end{description}
In the first case, we classify the equation as being partially integrable 
while in the second case as being nonintegrable.  
We  provide methods for finding
particular closed form solutions and illustrate them on some examples.  
In each case one looks for solutions related by a rational transformation
to the general solutions of first order nonlinear ODEs like Riccati, 
\index{Riccati equation}
Weierstrass or Jacobi possessing the PP, 
or solutions related by a nonlinear transformation like the logarithmic
derivative to a second or third order linear system with constant 
coefficients.

% ---------------------------------------------------------------------------
\section {Partially integrable equations}
\indent

\subsection{KPP equation}
\indent

\index{KPP equation}
The Kolmogorov, Petrovskii and Piskunov equation (KPP)
\cite{KPP,FitzHugh,NewellWhitehead}
%\refe{28} Nagumo {\it et al.} 1962 is not the correct reference
% {\rm (KPP)}\ E \equiv 
%u_t + p_i u_{xx} + q_i u^3 - \gamma u = 0,\ 
%p_i q_i \gamma \ne 0,\ u \in {\cal R}.

\begin{equation}
 E \equiv u_t - u_{xx} + {2\over d^2} (u-e_1)(u-e_2)(u-e_3) = 0,\ 
 e_j \hbox{ distinct},
\end{equation}
possesses two opposite families of singularities
\begin{eqnarray}
& &
u= d \chi^{-1} + s_1/3  -d C/6 + O(\chi),\ 
\hbox{indices~: }-1,4,\ 
{\cal D}=d\partial_x 
\\ 
& &s_1=e_1 + e_2 + e_3,\
   s_2=e_2 e_3 + e_3 e_1 + e_1 e_2,\
   s_3=e_1 e_2 e_3. 
\end{eqnarray}
with the condition on the SM coming from the index 4
\begin{equation}
Q_4 \equiv C [ -3 ( C_t + C C_x) + \prod_{k=1}^3 ( C + s_1 - 3 e_k) ] = 0
\end{equation}
Denoting $(j,l,m)$ any permutation of $(1,2,3)$ and $k_i = (3 e_i - s_1)/(3 d)$,
the one-family truncation 
\index{Weiss truncation}
$ u= d \chi^{-1} + s_1/3  -d C/6 $ yields the moving front solution~:
\begin{eqnarray}
 u  & = & 
       {e_l + e_m \over 2}
    + d {k \over 2} \tanh {k \over 2}(x-ct-x_0) 
\label{eqKPPFront}
\\
 k^2 & = & (k_l - k_m)^2,\ c= -3  (k_l + k_m)/2 .
\end{eqnarray} 
The two--family truncation like for MKdV or sG yields the stationary pulse
\begin{equation}
 u= e_1 +  {e_2 - e_3 \over \sqrt 2}
\sech i{e_2 - e_3 \over d \sqrt 2}(x-x_0),\ 
2 e_1 - e_2 - e_3=0. 
\label{eqStationaryPulse}
\end{equation}
A third very interesting solution representing  the collision of two fronts
\cite{KawaharaTanaka}
with different velocities can be easily found \cite{ConteMusette1993} with the
assumption
\begin{equation}
 u = {s_1 \over 3} + u_0 \partial_x \Log \psi,
\label{eqKPP2f}
\end{equation} 
where $\psi$ is the general solution of a linear system with constant 
coefficients
\index{Lax pair of third order}
\begin{eqnarray}
\hspace{-15 truemm}
& & \psi  =  \sum_{n=1}^3 C_n \exp\left[k_n (x+b_2t) +k_n^2 b_1 t \right],
\ C_n \hbox{ arbitrary,}\ C_1 C_2 C_3 \not=0    \\
\hspace{-15 truemm}
& &
\psi_{xxx} - a_1 \psi_x - a_2 \psi=0,\ \psi_{t} - b_1 \psi_{xx} - b_2 \psi_x=0
\end{eqnarray} 
with the following values of the coefficients and the constant $k_n$ 
\begin{eqnarray}
& &  
a_1   = (s_1^2 - 3 s_2 )/ (3 d^2),\  
a_2=(2 s_1^3 - 9 s_1 s_2 + 27 s_3) / (27 d^3),
\nonumber
\\
& &
b_1=-3,\ b_2=0,\
k_n  = (3 e_n - s_1)/ (3 d) .
\end{eqnarray}

\section{Nonintegrable equations}
\indent

\subsection{Kuramoto-Sivashinsky equation}
\indent

The Kuramoto-Sivashinsky (KS) equation
\index{Kuramoto-Sivashinsky (KS) equation}
\begin{equation}
\label{eqKScons} 
 E  \equiv 
 u_t + u u_x +  u_{xx} +  u_{xxxx}=0,
\end{equation}
describes, for instance, the fluctuation of the position of a flame front, 
or the motion of a fluid going down a vertical wall,
or a spatially uniform oscillating chemical reaction in a homogeneous medium. 
For a review, see ref.~\cite{Manneville1988}.

\noindent 
This equation possesses only one family of singularities 
\cite{ConteMusette1989} 
\begin{eqnarray}
& & u = 120 \chi^{-3} + 60 ( S+ 1/19) \chi^{-1} + ( C -15 S_x) + O(\chi),  \\
& & \hbox{indices~: } -1,6, (13\pm i \sqrt{71})/2,\ 
{\cal D} = 60{\partial_x}^3 + (
60/19)\partial_x
\end{eqnarray}
Due to the existence of the two complex irrational Fuchs indices the Laurent 
series depends
only on two arbitrary functions ($u_6$ and the arbitrary function in the
expansion variable $\chi$) whatever be the SM. The WTC truncation 
\index{Weiss truncation}
\begin{equation}
 u = 120 \chi^{-3} + 60 ( S + 1/19) \chi^{-1} + ( C - 15 S_x)
\end{equation}
generates three determining equations $ E_4=0,\  E_5=0,\ E_7 = 0 $ 
whose general solution  is 
$ C= \hbox{arbitrary }c,\  S= -11/38,1/38$.
This corresponds to the well-known travelling wave solution of Kuramoto and
Tsuzuki \cite{KuramotoTsuzuki} only existing for two values 
$k^2 = -1/19$ or $11/19 $~:
\begin{equation}
 u  =  c
   + \left({30 \over 19} k - 15 k^3 \right) \tanh {k \over 2}(\xi-x_0)
   + 15  k^3 \tanh^3 {k \over 2}(\xi-x_0)
\end{equation}
where $\xi = x-ct$ and $c, x_0$ are arbitrary constants.

\noindent This solution can also be retrieved with the assumption that $
u = c + {\cal D} \Log\psi$ with $\psi$ the general solution of a linear system
with constant coefficients
\index{Lax pair of second order}
\begin{equation}
\label{eqLinearKS} 
\psi_{xx} - (k^2/4)\psi = 0\ ,\ \psi_t - c \psi_x = 0.
\end{equation}

\noindent  Let us remark that the reduction $u(x,t) \to c + U(x-ct) $ of
the PDE (\ref{eqKScons}) yields the nonintegrable ODE 
\begin{equation}
 U''' + U' + U^2/2 + K = 0,\ K \hbox{ arbitrary}
\end{equation}
for which we have found, in the case $ K= -450 k^2/19^2 $, 
a Riccati sub-equation \index{Riccati equation} linearisable into the system (\ref{eqLinearKS}). 
The challenge not yet solved is to find for \underbar{every} $ K $ 
a closed form particular solution depending on one arbitrary constant.

Let us also mention the interesting work of Porubov 
\cite{Porubov1993,Porubov1996} who has found for a large class of nonlinear 
PDEs like
\begin{equation}
\label{eqBenardMarangoni}
\eta_t +a_1 \eta_x +a_2 \eta \eta_x +a_3 (\eta \eta_x)_x +a_4 \eta_{xx}
 +a_5\eta_{xxx} +a_6\eta_{xxxx} = 0
\end{equation} 
and for some particular values of the constant parameters $\{a_i\}$,
travelling wave solutions in terms of Weierstrass  elliptic functions or its
logarithmic derivative.
\index{elliptic function}

%The assumption (\ref{eqKSRiccati}) is equivalent to the Weiss truncation 
%method for $\chi=V^{-1}$ \index{Weiss truncation} \index{Riccati equation}
%and a Riccati system  (\ref{eqChiX})--(\ref{eqChiT}) with constant 
%coefficients $(S,C)$\cite{FournierSpiegelThual,ConteMusette1989}.

%The fourth order ODE (\ref{eqKSReduc1}) admits the exact solution
% (\ref{eqKSsolKT}) with $c=0$ ($c$ only reflects the galilean invariance), 
%depending on only one arbitrary parameter $x_0$.
%This is far away from the four arbitrary parameters one could expect. 
%However, the chaotic nature of KS equation forbids the existence of a 
%four-parameter solution to (\ref{eqKSReduc1}), and more
%precisely the two irrational Fuchs indices $(13 \pm i \sqrt{71}) / 2$
%\cite{FournierSpiegelThual} reduce by two the number of available arbitrary 
%parameters. 
%In other words, there probably exists a two-parameter closed form solution 
%to the ODE (\ref{eqKSReduc1}), but it has not yet been found.

%{\it Example KS}. In the third order ODE for $U$,  
%one assumes the existence of a particular
%solution $U$ polynomial in  the solution $V$ of a Riccati equation
%\cite{AiraultKaliappan} % Airault and Kaliappan
%\begin{equation}
%\label{eqKSRiccati}
% U=\sum_{j=0}^3 a_j V^j,\ {dV \over d\xi}+V^2 -{k^2 \over 4}=0
%\end{equation} 
%and one identifies the LHS of eq.~(\ref{eqKScons}) to the null polynomial
% in $V$;
%this generates seven equations in the five unknowns $a_j,k^2$ and this yields
% the solution
%(\ref{eqKSsolKT})  depending on two arbitrary parameters $(c,x_0)$.
% 

For the KS equation with an additional dispersive term   
\begin{equation}
\label{eqKSdispersive} 
E  \equiv 
 u_t + u u_x +  u_{xx} + b u_{xxx} + u_{xxxx}=0,
\end{equation}
which belongs to the previous class (\ref{eqBenardMarangoni}), 
Kudryashov \cite{Kudryashov} has found a particular solution 
depending on three
arbitrary constants in terms of the Weierstrass elliptic function
\index{elliptic function}
 and its derivative in the
single case $b^2=16$. 
An easy way to find it is to assume the existence of a
solution  of the form
\begin{equation}
 u  = 
 a_0 + a_2 \wp(x-ct-x_0,g_2,g_3) + a_3 \wp'(x-ct-x_0,g_2,g_3),
\end{equation}
compatible with the singularity structure of its Laurent series expansion.
 Enforcing in the LHS $E$ the conditions
$\wp'^2=4 \wp^3 - g_2 \wp - g_3$ and
$\wp''=6 \wp^2 - (g_2 / 2)$, 
one identifies to zero a polynomial of two variables $(\wp,\wp')$ of
degree one in $\wp'$. 
This similarly generates six equations in the five unknowns $a_j,g_2,g_3$
and yields a nondegenerate elliptic solution only if $ b^2 = 16 $ and
\begin{equation} 
{\hskip -2 truemm}
 a_0=c - 4 / b,\ a_2=- 15 b,\ a_3= - 60,\ g_2=1/12,\ 
(g_3,c,x_0) \hbox{ arbitrary}.
\end{equation}

\subsection{Complex Ginzburg-Landau equation CGL3}
\indent

The one-dimensional cubic complex Ginzburg-Landau (CGL3) equation
\index{complex Ginzburg-Landau (CGL3) equation}
\begin{equation}
\label{eqCGL}
 E \equiv i u_t + p u_{xx} + q \vert u \vert^2 u - i \gamma u = 0,
\ p q \ne 0,\ (u, p, q) \in {\cal C},\ \gamma \in {\cal R}, 
\end{equation} with $p,q,\gamma$ constants,
describes pattern formation and coherent structures in many different domains~:
Taylor-Couette flows between coaxial rotating cylinders, wave propagation in
optical fibers and chemical reactions. 
For a review see \cite{CrossHohenberg1993}.
This PDE is physically strongly connected to the KS equation and it also
possesses two irrational complex conjugate indices. 
As is the case for the AKNS system\index{AKNS system},
CGL3 equation possesses four families of singularities
\begin{eqnarray}
\hspace{-18.5 truemm}
& &u =  A_0 \chi^{-1+i \alpha} ( 1 + A_1 \chi + O(\chi^2)),\  
   \overline{u} =  B_0 \chi^{-1 - i \alpha} ( 1 + B_1 \chi + O(\chi^2)),
\\
\hspace{-18.5 truemm}
& & A_0 B_0 = 3 \vert{p^2}\vert\alpha / D_i,\ 
\alpha = D_r\pm\sqrt{D_r^2  + 8 D_i^2/9}, D_i= p_r q _i - p_i q_r,
\\
\hspace{-18.5 truemm}
& &\  D_r = (p_r q _r + p_i q_i),\ 
\hbox{Fuchs indices~: }-1,0,7/2 \pm \sqrt{1-24 \alpha^2}/2.
\end{eqnarray}
The important information we get from the singularity analysis is that neither
$(u,\overline{u})$, nor ($\vert u\vert,\arg u$) nor (Re $u$,Im$u$) have a simple
singularity structure. 
In this framework, better variables are $(Z,\theta)$ defined as
\begin{equation}
u = Z e^{i \theta},\ \theta=\alpha\Log \psi + \theta_0,\ \psi/\psi_x = \chi,
\end{equation}
where $\theta_0$ is an arbitrary function representing the index $0$. 
Then $Z$ and grad$\theta$ behave like simple poles, so that the usual methods
are applicable. 
\index{Lax pair of second order}
\index{Lax pair of third order}
With $\psi$ the general solution of a second order or third order system and
with a one--family or two--family truncation for $(Z,\theta)$ one finds
\cite{ConteMusette1993}
\underbar{all} the known closed form solutions of this equation. 
Among them, a very interesting solution representing 
a ``collision of two shocks'' \cite{NozakiBekki1983,NozakiBekki1984}
\index{Lax pair of third order}
\begin{eqnarray}
& &
u = A_0 
  {k \over 2} 
 {\sinh k x/2 \over \cosh k x/2 + e^{-3 \gamma (t - t_0)/2}}
 e^{i[\expon \Log(1 + e^{3 \gamma (t-t_0)/2} \cosh k x/2)]},
%\\ A_0^2 & = &
% {9 \over 2} {\pcarre \over \pqi^2} (\pqr \pm \Delta),\
%   \expon        ={3 \over 2                 \pqi  } (\pqr \pm \Delta)
%                 ={\pqi \over 3 \pcarre} \GLA_0 \GLB_0
%\\
%\Delta & = &
%\sqrt{\pqr^2 + {8 \over 9} \pqi^2},\
%   \pqr=p_r q_r + p_i q_i,\ \pqi=p_r q_i - p_i q_r,
\nonumber
\\
& &
\ k^2  =  - 2 \gamma / p_i,\ p_r=0,
\label{eqCollision}
\end{eqnarray}
is easily found by assuming that $\psi$ satisfies a third order linear system.
For $q_r=0$ (hence $\alpha =0$) this solution degenerates to the 
``collision of two fronts'' solution of KPP previously considered.
\index{KPP equation}
The Table \ref{TableCGL3} summarises the known solutions of CGL3 and their
degeneracies to NLS and KPP equations.
An open problem is to find the solution of CGL3 which degenerates for
$p_i=q_i=\gamma=0$ to the bright soliton (\ref{eqBright}) of NLS.

\begin{table}[h]
\caption{Degeneracies of the known solutions of CGL3.
}\label{TableCGL3}
\vspace{0.4cm}
\begin{tabular}{| l | l | l |}
\hline CGL3 & NLS $(p_i=q_i=\gamma=0)$ & KPP $(p_r=q_r=0)$
\\ \hline \hline
propagating hole \cite{BekkiNozaki1985,ConteMusette1993}
&
dark soliton (\ref{eqDark})
&
\\ \hline
shock or front \cite{NozakiBekki1984,ConteMusette1993}
&
&
front (\ref{eqKPPFront})
\\ \hline
pulse or solitary wave \cite{PS1977,ConteMusette1993}
&
&
stationary pulse (\ref{eqStationaryPulse})
\\ \hline
$
\displaystyle
\hbox{collision of two shocks }
 (p_r=0)
\atop
\displaystyle
(\ref{eqCollision})
\cite{NozakiBekki1983,NozakiBekki1984,ConteMusette1993}
$
&
&
collision of two fronts (\ref{eqKPP2f})
\\
\hline
\end{tabular}
\end{table}

\begin{theindex}                     % Orme 10pt
%input Cargese96.ind
%input Musette10.idx                   % Orme 10pt
\end{theindex}                       % Orme 10pt

\printindex

% ***************************************************************** References
%\section{References} % in alphabetical order
%\pagestyle{myheadings}
%\markboth{{\rm M.Musette \hspace{5cm}}}{{\rm \hspace{2.5cm} Nonlinear partial 
%differential equations}}
 %{99}
\vfill \eject
\end{document}